\documentclass[twocolumn,groupedaddress,assymb,amsmath]{revtex4}
\usepackage{}
\usepackage{graphicx}
\usepackage{amssymb}
\usepackage{amsmath}

\usepackage{extarrows}
\usepackage{amsfonts}
\usepackage{epstopdf}
\usepackage{mathrsfs}
\usepackage[bookmarks=false]{hyperref}
\hypersetup{colorlinks=true,citecolor=blue,linkcolor=blue,urlcolor=blue,pdfstartview=FitH,bookmarksopen=true}

\usepackage{color}
\definecolor{MyDarkBlue}{rgb}{0,0.08,0.45}
\definecolor{yellow}{rgb}{0.99,0.99,0.70}
\definecolor{white}{rgb}{1.0,1.0,1.0}
\definecolor{black}{rgb}{0.00,0.00,0.00}
\definecolor{green}{rgb}{0.8,0.98,0.83}
\pagecolor{green}
\definecolor{zzz}{rgb}{0.9,0.0,0.4}

\begin{document}
\title{Photon blockade in non-Hermitian optomechanical systems with nonreciprocal couplings}
\author{ J. Y. Sun,$^{1}$ and H. Z. Shen,$^{1,2,}$\footnote{ \textcolor{zzz}{Corresponding author: shenhz458@nenu.edu.cn }}}
\affiliation{$^1$Center for Quantum Sciences and School of Physics,
Northeast Normal University, Changchun 130024, China\\
$^2$Center for Advanced Optoelectronic Functional  Materials
Research, and Key Laboratory for UV Light-Emitting Materials and
Technology of Ministry of Education, Northeast Normal  University,
Changchun 130024, China}
\date{\today}

\begin{abstract}
We study the photon blockade at exceptional points for a non-Hermitian optomechanical system coupled to the driven whispering-gallery-mode microresonator with two nanoparticles under the weak optomechanical coupling approximation, where exceptional points emerge periodically by controlling the relative angle of the nanoparticles. We find that conventional photon blockade occurs at exceptional points for the eigenenergy resonance of the single-excitation subspace driven by a laser field, and discuss the physical origin of conventional photon blockade. Under the weak driving condition, we analyze the influences of the different parameters on conventional photon blockade.
We investigate conventional photon blockade at non-exceptional points, which exists at two optimal detunings due to the eigenstates in the single-excitation subspace splitting from one (coalescence) at exceptional points to two at non-exceptional points.
\textbf{Unconventional photon blockade can occur at non-exceptional points, while it does not exist at exceptional points since the destructive quantum interference cannot occur due to the two different quantum pathways to the two-photon state being not formed.}
The realization of photon blockade in our proposal provides a viable and flexible way for the preparation of single-photon sources in the non-Hermitian optomechanical system.
\end{abstract}


\maketitle
\section{Introduction}

Photon blockade (PB) and tunneling are current research topics and play an essential role in various fundamental studies and practical applications \cite{Bajcsy15025014,Bamba99171111,Carusotto85299,Ferretti82013841,Ferretti15025012,Ferretti85033303,Flayac88033836,
Gerace89031803,Greentree2252,Gheri602673,Imamoglu791467,Kyriienko90063805,Majumdar21160,Majumdar87235319,Rebic65063804,
Zhou92023838,Giovannetti222,Knill46,Kok135,Li053837,Wang053832,Shen023805,Ren053710,Stefanatos013716}. In these pioneering studies, PB is generated in weakly nonlinear systems, allowing for destructive quantum interference between distinct driven-dissipative pathways \cite{Liew104183601,Bamba83021802,Flayac96053810, Carmichael2790,Casalengua1900279,Kyriienko033807}, called unconventional PB (UPB).
Based on this fundamental principle, many quantum systems are predicted to have PB effect with weak nonlinearities, such as the nonlinear photonic molecule \cite{Xu90043822,Xu033832,Xu033809}, optical cavity with a quantum dot \cite{Zhang043832,Tang9252,Liang053713,Wang64003}, coupled single-mode cavities with second- or third-order nonlinearity \cite{Flayac013815,Flayac11223,Shen32835,Shen035503,Zhou17332,Zou16175,Roberts021022}, coupled optomechanical system \cite{Xu035502,Savona5937,Qu043823,Hou063817}, a gain cavity \cite{Zhou043819}, exciting polaritons \cite{Carreno196402}, non-Markovian system \cite{shen023856,Shen013826}, and Gaussian squeezed states \cite{Lemonde90063824,Sarma053827}.
On the other hand, PB arises from the anharmonicity in the eigenenergy of the systems caused by strong nonlinearity \cite{Huang121153601,Birnbaum87,Shen023849,Zhou033713,Lee013834,Kim397500,Smolyaninov187402,Zheng107223601,Majumdar183601}, called conventional PB (CPB). There are various systems for producing CPB, such as cavity quantum electrodynamics (QED) systems \cite{Hennrich94053604,Hamsen118133604,Ridolfo109193602,Tian6801,Werner011801,Brecha2392,Rebic490,Liang063834,Yan5086,Fink011012,Miranowicz023809,
Zhu063842,Faraon033838,Liao053836,Lin053850}, circuit QED systems \cite{Lang106243601,Hoffman107053602,Liu89043818,Leib093031}, optomechanical systems \cite{Rabl107063601,Wang92033806,Zhai07654,
Nunnenkamp063602,Liao023853,Komar013839,Lu2943,Xu013818,Miranowicz013808,Xie063860,Xie013861,Zou043837,Liu032101},
coupled cavities \cite{Angelakis76031805,Shen91063808}, a two-level system coupled to the cavity \cite{Grujic103025,Boite033827,Radulaski011801,Li043724,Xue4424,Jing033707,Yao054004}, dynamical blockade \cite{Ghosh013602}, a quantum dot in a photonic crystal system \cite{Faraon859}, and a quantum dot coupled to a nanophotonic waveguide \cite{Foster173603}.

Experimentally \cite{Vaneph043602,Snijders043601}, the signature of PB and tunneling can be distinguished by measuring the second-order correlation function $g^{(2)}(0)$ \cite{Loudon2000}. For the PB (or the photon antibunching $g^{(2)}(0)<1$), driven by an external coherent field, the presence of a single photon in a system will hinder the coupling of the subsequent photons because of the strong nonlinearities presenting in the quantum system, while for the photon tunneling (or the photon bunching $g^{(2)}(0)>1$), the coupling of initial photons will favor the coupling of the subsequent photons \cite{Loudon2000}.
The potential applications of PB include the realizations of interferometers \cite{Gerace281}, quantum nonreciprocity \cite{Fratini243601,Mascarenhas54003}, and single-photon transistors \cite{Chang807}.

The nonlinear interaction between optical and mechanical modes arising from the radiation pressure force in cavity optomechanical (COM) systems exhibits many interesting nonlinear effects such as photon (phonon) blockade \cite{Rabl107063601,Ramos110193602,Wang093006}, nonreciprocity \cite{Xu143,Manipatruni213903,Shen657,Bernier604,Jiang064037,Li630,Liu15382,Shang115202,Gao25161,Xu053854,Mirza25515,Jiao143605}, optomechanical-induced transparency \cite{Agarwal81041803,Teufel471204,Zhang025199}, and nonlinearity \cite{Aldana88043826,Zhou88063854,Lu093602}. Cavity optomechanics has received significant attention in both fundamental experiments \cite{Teufel475359,Chan47889} and sensing applications \cite{Abbott11073032,Pontin89023848}. Currently, experimental techniques of cavity optomechanics are still in the single-photon weak-coupling regime \cite{Aspelmeyer861391}. However, to date, only a few realizations like cold-atomic clouds in the optomechanical cavity \cite{Brennecke322235,Gupta99213601} have met the requirements.

In parallel, properties and applications of non-Hermitian systems \cite{Sergi1350163,Zloshchastiev1298,Sergi062108}, especially exceptional points (EPs) systems, have attracted intense interest in recent years \cite{Bender805243,Bender70947,Konotop88035002,Feng11752,Peng10394,Ruter6192,Zyablovsky571063,Longhi12064001,Lu014006}. In such systems, two or more eigenstates coalesce at EPs, leading to a variety of unconventional effects observed in experiments, such as loss-induced coherence \cite{Guo103093902,Peng346328}, PB induced by dissipation and chirality \cite{Zuo043715,Jing09492}, unidirectional lasing \cite{Longhi5B1}, unidirectional invisibility \cite{Lin106213901}, robust wireless power transfer \cite{Assawaworrarit546387}, and exotic topological states \cite{Zhou3591009,Yang3591013}. The EP effects in COM have also been probed both theoretically and experimentally \cite{Jing113053604,Xu53780,Lu8044020,Jing73386}, such as the low-power phonon laser \cite{Jing113053604,Lu8044020}, high-order EPs in COM \cite{Jing73386}, and nonreciprocal COM devices \cite{Xu53780}, highlighting new opportunities for enhancing or steering coherent light-matter interactions using the new tool of EPs.
Very recently, by coupling a whispering-gallery-mode (WGM) microresonator with two external nanoparticles, the periodic emergence of EPs has been observed experimentally for tuning the relative positions of the particles \cite{Peng1136845}. The counterintuitive EP effects, such as modal chirality \cite{Peng1136845,Cao033907} and highly sensitive sensing \cite{Chen548192,Tian043810}, have been revealed in these exquisite devices.

However, we find that the connections between the PB and EPs have not been studied in the non-Hermitian optomechanical system. To be specific, the key to addressing the problem is to explore whether and how EPs affect PB.

In this paper, we investigate the PB effects at EPs in a non-Hermitian optomechanical system coupled to the driven WGM microresonator with two nanoparticles, where the coupling between clockwise- (CW) and counterclockwise- (CCW) traveling waves is nonreciprocal. By tuning the relative position of two nanoparticles, the system can be steered to an EP or away from it, where the photon statistical properties are well controlled such that CPB at EPs is realized. Moreover, we discuss the origin of CPB at EPs.

Our scheme has the following features. (i) CPB occurs at EPs for the eigenenergy resonance of the single-excitation subspace driven by a laser field. (ii) CPB can be found at non-EPs for two optimal detunings. \textbf{(iii) UPB at EPs does not occur due to the two different quantum interference pathways being not formed, but it can exist at non-EPs.}

The remainder of the paper is organized as follows. In Sec.~II, the theoretical model and Hamiltonian are described for the WGM microresonator coupled with mechanical mode. In Sec.~III, the photon statistical properties and EPs of the system are discussed. The physical origin of CPB at EPs is revealed. In Sec.~IV, we give the analytical solution of the second-order correlation function and study the influences of the different parameters on CPB. Moreover, CPB at non-EPs is discussed. \textbf{In Sec.~V, we investigate UPB at EPs and non-EPs.} Finally, the main results are summarized in Sec.~VI.

\section{Model and Hamiltonian}
  \begin{figure}[t]
   \centerline{
   \includegraphics[width=8.5cm, height=6.5cm, clip]{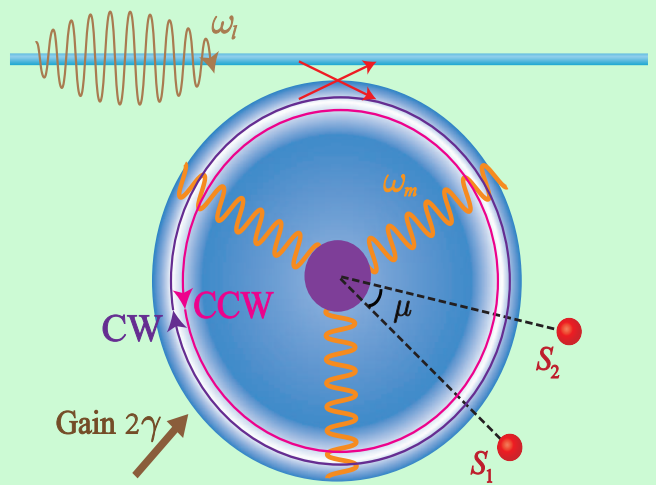}}
   \caption{Optomechanics in a microresonator is perturbed by two nanoparticles in the WGM field. The waveguide is driven by a laser with the frequency $\omega_l$ to the WGM microresonator through tapered fiber. The resonator with the cavity effective gain rate $2\gamma$ supports a mechanical mode at the frequency $\omega_m$. $\mu$ is the relative angle between the two particles denoted by $S_1$ and $S_2$. By tuning the relative phase angle $\mu$ between the particles, one can control the nonreciprocal couplings $\mathcal {E}_{1}$ and $\mathcal {E}_{2}$ given by Eq.~(\ref{nonreciprocal}) between CW and CCW modes, which results in periodical revival and suppression of mode splitting and coalesce.}\label{device}
 \end{figure}
As depicted in Fig.~\ref{device}, we consider a WGM resonator consisting of two optical modes, where the coupling between two modes is nonreciprocal, which can be achieved by two nanoparticles \cite{Zhu1823535}. This resonator driven by a laser with the frequency $\omega_l$ also supports a phonon mode at the mechanical frequency $\omega_m$.
Two silica nanotips as Rayleigh scatterers are placed in the evanescent field of the resonator, which are fabricated by the wet etching of tapered fiber tips prepared by heating and stretching standard optical fibers. The position of each particle is controlled by a nanopositioner, which tunes the relative position and effective size of the nanoparticle in the WGM fields. The non-Hermitian optical coupling of the CW and CCW traveling waves induced by the nanoparticles is described by the scattering rates \cite{Wiersig063828,Wiersig203901,Li1909}
 \begin{equation}
  \begin{aligned}
   {\mathcal {E}_{1}} &= {\lambda _1} + {\lambda _2}{e^{  i2m\mu }},\\
   {\mathcal {E}_{2}} &= {\lambda _1} + {\lambda _2}{e^{ - i2m\mu }},\label{nonreciprocal}
  \end{aligned}
 \end{equation}
where $\mathcal {E}_{1}$ ($\mathcal {E}_{2}$) corresponds to the scattering from the CCW (CW) mode to the CW (CCW) mode. $m$ is the azimuthal mode number. $\mu$ denotes the relative angular position of the two scatterers. $\lambda_j$ $(j=1,2)$ is the complex frequency splitting induced by the $j$th scatterer, which depends on the volume of the $j$th particle within the WGM fields, and is tuned by controlling the distance between the particle and resonator with nanopositioners. Steering the angle $\mu$ can bring the system to EPs, as already observed experimentally \cite{Chen548192,Peng1136845}. \textbf{Moreover, we present some discussions on the experimental implementation of Eq.~(\ref{nonreciprocal}) in Appendix \ref{experimental}.}
In the rotating frame ${\hat{V}}_{1} = \exp [ - i{\omega _l}t(\hat{a}_1^\dag {\hat{a}_1} + \hat{a}_2^\dag {\hat{a}_2})]$, the total non-Hermitian Hamiltonian of the system is written as ($\hbar \equiv 1$)

\begin{equation}
   \begin{aligned}
      {\hat{H}_{T}} = &\Delta_1 {\hat{a}_1^\dag} {\hat{a}_1} + \Delta_2 {\hat{a}_2^\dag} {\hat{a}_2} +{\omega _{m}}{\hat{b}^\dag }\hat{b} + {\mathcal {E}_1}{\hat{a}_1^\dag} {\hat{a}_2} + {\mathcal {E}_2}{\hat{a}_2^\dag} {\hat{a}_1}\\
              &- g({\hat{b}^\dag } + \hat{b})(\hat{a}_1^\dag {\hat{a}_1} + \hat{a}_2^\dag {\hat{a}_2})+F{\hat{a}_1^\dag}  + {F }{\hat{a}_1},
\label{hamiltonian3}
   \end{aligned}
 \end{equation}
where $\hat{a}_1$ ($\hat{a}_2$) and $\hat{a}_1^\dagger$ ($\hat{a}_2^\dagger$) denote the photon annihilation and creation operators of the CW (CCW) mode, respectively, satisfying $[\hat{a}_1,\hat{a}_1^\dag]=1$ and $[\hat{a}_2,\hat{a}_2^\dag]=1$. $\hat{b}$ ($\hat{b}^\dag$) is the phonon annihilation (creation) operator of the mechanical mode. $\Delta_j=\omega_j-\omega_l$ is the detuning between cavity and laser with $\omega_j=\omega_0 - i {\gamma_j}/2 +\lambda_1+\lambda_2$, where $\omega_0$ is the frequency of the bare system. The effective loss rate $\gamma_j=\gamma_j^{i}-\xi$ is reduced by the gain $\xi$ (round-trip energy gain) and intrinsic loss rate $\gamma_j^{i}$ ($\gamma_j<0$) \cite{McCall289,Armani925,Peng394}. In consideration of a small change in the cavity length, the cavity optomechanical coupling coefficient is written as $g=\omega_0/\sqrt{2 R^{2} m_{\rm{eff}} \omega_m}$, where $R$ is the radius of the resonator, and $m_{\rm{eff}}$ denotes the effective mass of the mechanical mode. $ F = \sqrt {2{| \gamma_1 | }P/(\hbar {\omega _l})}$ denotes the amplitude of the laser field with power $P$.
\begin{figure*}[t]
  \centerline{
  \includegraphics[width=13.5cm, height=6.2cm, clip]{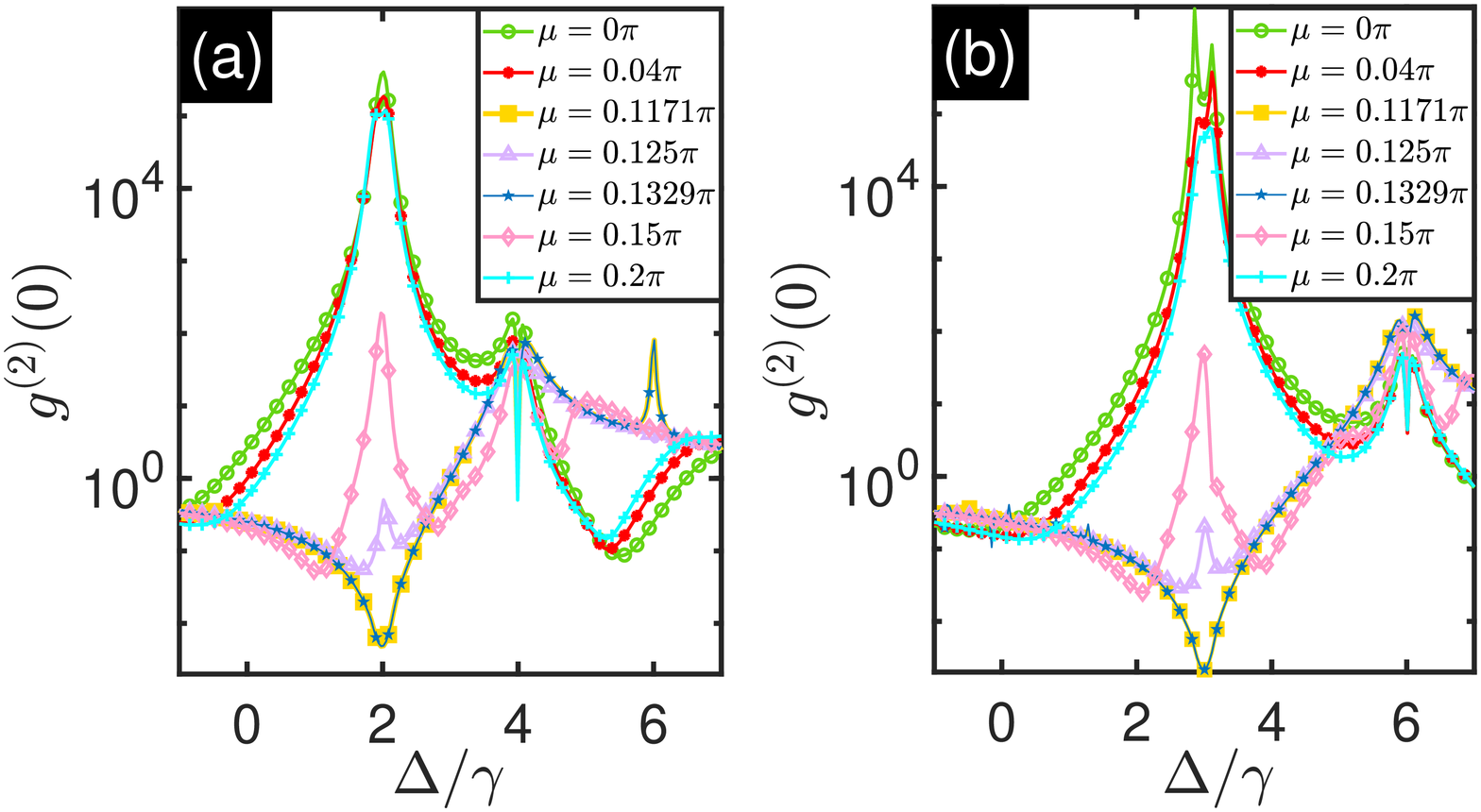}}
  \caption{The second-order correlation function $g^{(2)}(0)$ as a function of the detuning $\Delta$ with various relative angular positions $\mu$ of two nanoparticles is solved by master equation~({\ref{masterequation}}). The parameters chosen are $\lambda_1=(1.5-0.355i)\gamma$, $\lambda_2=(1.4-0.645i)\gamma$, $m=4$, $F=0.1\gamma$, (a) $U=2\gamma$; (b) $U=3\gamma$.}\label{figure1}
\end{figure*}

 With the unitary transformation ${\hat{V}}_{2} = \exp [{g}/{\omega _m}(\hat{a}_1^\dag {\hat{a}_1} + \hat{a}_2^\dag {\hat{a}_2})({\hat{b}^\dag}- \hat{b})]$, Eq.~(\ref{hamiltonian3}) becomes
 \begin{equation}
  \begin{aligned}
   {\hat{H}_{\rm{ker}}} =& \Delta_1 {\hat{a}_1^\dag} {\hat{a}_1} + \Delta_2 {\hat{a}_2^\dag} {\hat{a}_2} + {\omega _{m}}{\hat{b}^\dag }\hat{b} + {\mathcal {E}_1}{\hat{a}_1^\dag} {\hat{a}_2} + {\mathcal {E}_2}{\hat{a}_2^\dag} {\hat{a}_1}\\
    &-{{g}^2}/{\omega _{m}}[(\hat{a}_1^\dag {\hat{a}_1})^2 + (\hat{a}_2^\dag {\hat{a}_2})^2 + 2\hat{a}_1^\dag {\hat{a}_1}\hat{a}_2^\dag {\hat{a}_2}]\\
    &+F{\hat{a}_1^\dag}  + {F}{\hat{a}_1}, \label{hamiltonian4}
   \end{aligned}
 \end{equation}
 where we made the weak optomechanical coupling approximation, i.e., ${g}/{\omega _m} \ll 1$. The derivation details of Eq.~(\ref{hamiltonian4}) can be found in Appendix \ref{DerivationHamiltonian}. We find that the mechanical mode is decoupled with the optical cavity, which means the evolutions of optical and mechanical parts are independent of each other, i.e., the state evolution of the total system ${e^{ - i{\hat{H}_{\rm{ker}}}t}}{| \psi  \rangle _{sys}} = {e^{ - i{\hat{H}_{\rm{eff}}}t}}{| \psi \rangle _{opt}} \otimes {e^{ - i{\omega _m}{\hat{b}^\dag }\hat{b}t}}{| \psi \rangle _{mech}}$. When we study the photon statistical properties in the system, the mechanical part in Eq.~(\ref{hamiltonian4}) can be ignored safely, and then Eq.~(\ref{hamiltonian4}) becomes
 \begin{equation}
\begin{aligned}
    {\hat{H}_{\rm{eff}}}= &\Delta_1 \hat{a}_1^\dag {\hat{a}_1} + \Delta_2 \hat{a}_2^\dag {\hat{a}_2} + {\mathcal {E}_1}\hat{a}_1^\dag {\hat{a}_2} + {\mathcal {E}_2}\hat{a}_2^\dag {\hat{a}_1} \\
               &- U(\hat{a}_1^\dag {\hat{a}_1}\hat{a}_1^\dag {\hat{a}_1} + \hat{a}_2^\dag {\hat{a}_2}\hat{a}_2^\dag {\hat{a}_2} + 2\hat{a}_1^\dag {\hat{a}_1}\hat{a}_2^\dag {\hat{a}_2}) \\
               &+ F\hat{a}_1^\dag  + {F}{\hat{a}_1},
\label{hamiltonian5}
\end{aligned}
 \end{equation}
where $U = {{{{g}^2}} \mathord{/  {\vphantom {{{{g'}^2}} {{\omega _m}}}} .  \kern-\nulldelimiterspace} {{\omega _m}}}$ denotes the Kerr-type nonlinear strength induced by the optomechanical coupling. \textbf{We also discuss the equivalence between the original Hamiltonian~(\ref{hamiltonian3}) and effective Hamiltonian~(\ref{hamiltonian5}) under the weak optomechanical coupling approximation in Appendix \ref{validity}, which shows that the used approximation is valid.}

The effective Hamiltonian ({\ref{hamiltonian5}}) without the effective loss rate can be partitioned into Hermitian and anti-Hermitian parts: $\hat{H}_{\rm{eff}}=\hat{H}_{+} + \hat{H}_{-}$, where we have $\hat{H}_{+}= \hat{H}_{+}^\dag$ and $\hat{H}_{-}=- \hat{H}_{-}^\dag$. To correctly account for the driven-dissipative character of the system, we introduce the Lindblad master equation for the system density matrix
  \begin{equation}
   \begin{aligned}
   \frac{{d{\rho} }}{{dt}} = &- i[ {{\hat{H}_ + },{\rho} } ] - i\{ {{\hat{H}_ - },{\rho} } \}+ \sum\limits_j {({\gamma _j}/2) \mathcal{D}({\rho} ,{\hat{a}_j})}\\
   & + 2i{\rm Tr}({\rho} {\hat{H}_- }){\rho},\label{masterequation}
   \end{aligned}
  \end{equation}
 where $\mathcal{D}({\rho} ,\hat{o}) = 2 \hat{o} {\rho} {{\hat{o}}^\dag } - {\hat{o}^\dag }\hat{o} {\rho}  - {\rho} {\hat{o}^\dag } \hat{o}$ denotes the Lindblad superoperator term for the annihilation operator $\hat{o}$ acting on the density matrix ${\rho}$ to account for losses to the environment. $\{ {{\hat{H}_ - },{\rho} } \}$ is defined as ${\hat{H}_ - }{\rho}+{\rho}{\hat{H}_ - }$. $\gamma _1$ and $\gamma _2$ denote the effective damping constants of CW and CCW modes, respectively. Without loss of generality, we assume that the decay rates and eigenfrequencies of the resonator modes are respectively equal, i.e., $|\gamma _1|=|\gamma _2|=2 \gamma$, $\omega_1=\omega_2=\omega$ ($\Delta _1=\Delta _2=\Delta \equiv \omega-\omega_l$).
 The steady-state solution $\rho _{s}$ of the density matrix ${\rho}$ is obtained by setting ${{d{\rho} } /{dt}} = 0$ in Eq.~(\ref{masterequation}).

\section{Photon statistical properties}
In this section, we analyze CPB at EPs in detail and investigate the photon statistical properties with various relative angular positions $\mu$ of two nanoparticles, which are carried out by simulating the quantum master equation numerically.
First of all, the photon statistical properties of CW mode are described by the second-order correlation function of the steady state defined by
 \begin{equation}
   \begin{aligned}
   g^{(2)}(0) = \frac{{\langle {\hat{a}_1^\dag \hat{a}_1^\dag {\hat{a}_1}{\hat{a}_1}} \rangle }}{{{{\langle {\hat{a}_1^\dag {\hat{a}_1}} \rangle }^2}}}
    = \frac{{{\rm Tr}({{\rho} _s}\hat{a}_1^\dag \hat{a}_1^\dag {\hat{a}_1}{\hat{a}_1})}}{{{{[{\rm Tr}({{\rho} _s}\hat{a}_1^\dag {\hat{a}_1})]}^2}}},
   \label{correlation_function}
   \end{aligned}
  \end{equation}
 which emphasizes the joint probability of detecting two photons at the same time. The case of $g^{(2)}(0) <  1$ $(g^{(2)}(0) > 1)$ corresponds to photon antibunching (bunching) in the cavity mode, which is a nonclassical effect.

Figure~{\ref{figure1}} shows $g^{(2)}(0)$ as a function of the detuning $\Delta$ with various angles $\mu$, which is solved by Eq.~({\ref{masterequation}}).
In order to exactly obtain the numerical results about detuning, we assume $\Delta$ to be real by choosing proper parameters \cite{Peng1136845,Vahala839,Spillane013817,Huet133902}: $\lambda_1=(1.5-0.355i)\gamma$, $\lambda_2=(1.4-0.645i)\gamma$, $m=4$, $F=0.1\gamma$. In Fig.~\ref{figure1}(a), the bunching ($g^{(2)}(0) > 1$) is observed at $\mu=0,0.04\pi,0.2\pi$ and the maximum bunching $g^{(2)}(0) \sim 10^5$ is obtained for $\Delta/\gamma=2$. Changing $\mu$ to $0.125\pi$ and $0.15\pi$, the bunching effects are obviously weakened, as shown the purple-triangle line and pink-diamond line in Fig.~\ref{figure1}(a).
However, at $\mu=0.1171\pi,0.1329\pi$, the most striking feature is the occurrence of PB ($g^{(2)}(0) \sim 0.005$) when the driving field is in resonance with the cavity, i.e., $\Delta=U=2\gamma$, as seen the blue-pentagram line and yellow-square line in Fig.~\ref{figure1}(a).
Moreover, increasing the Kerr-type nonlinear strength to $U=3\gamma$ has the similar observation with Fig.~\ref{figure1}(a) for $\Delta/\gamma=3$, as shown in Fig.~\ref{figure1}(b).
\textbf{Appendix {\ref{SupplementaryFigure}} presents the discussion of special points $\Delta=4\gamma$ and $\Delta=6\gamma$ in Fig.~{\ref{figure1}}(a) and (b).}

To gain more insights into CPB shown in Fig.~\ref{figure1}, we investigate the eigenenergy of the non-Hermitian Hamiltonian.
In the weak driving regime ($F \ll \gamma $), the Hilbert space of this system can be restricted in a subspace with few photons spanned by the basis states $\{ {| {n_1,n_2} \rangle | {N \le 2} } \}$ with the total excitation number $N = n_1 + n_2$, which denotes the Fock state with $n_1$ photons in the bare CW mode and $n_2$ photons in the CCW mode.
In the single-excitation subspace, we write the eigenenergies of the non-Hermitian Hamiltonian~({\ref{hamiltonian5}}) without the driving term as
\begin{equation}
\begin{aligned}
 E_1^ \pm  = \omega  - U  + c_1^\pm,
\label{E1}
\end{aligned}
 \end{equation}
with the corresponding non-normalized eigenstates
\begin{equation}
\begin{aligned}
 | {\psi _1^ \pm } \rangle = \pm \sqrt {{\mathcal {E}_2}} | {0,1} \rangle  + \sqrt {{\mathcal {E}_1}} | {1,0} \rangle,
\label{psi_1}
\end{aligned}
 \end{equation}
where $c_1^\pm =  \pm \sqrt {{\mathcal {E}_1}{\mathcal {E}_2}} $. Moreover, we also obtain the eigenenergies $E_2^s = 2\omega  - 4 U  + c_2^s$ and corresponding non-normalized eigenstates $| {\psi _2^\pm} \rangle = \sqrt 2 {\mathcal {E}_2}| {0,2} \rangle  + c_2^\pm| {1,1} \rangle  + \sqrt 2 {\mathcal {E}_1}|{2,0}\rangle $, $| {\psi _2^0} \rangle = {\mathcal {E}_2}| {0,2} \rangle  - {\mathcal {E}_1}|{2,0}\rangle $ in the two-excitation subspace, where $s=\pm,0$, ${c_2^\pm} =  \pm 2 \sqrt {{\mathcal {E}_1}{\mathcal {E}_2}} $, and ${c_2^0} = 0$. This shows that the eigenmode structure depending on the asymmetry of the coupling coefficients $\mathcal {E}_{1}$ and $\mathcal {E}_{2}$ can be tuned by controlling the relative angular position $\mu$ between the nanoparticles.

\begin{figure}[t]
  \centerline{
  \includegraphics[width=8.7cm, height=5.2cm, clip]{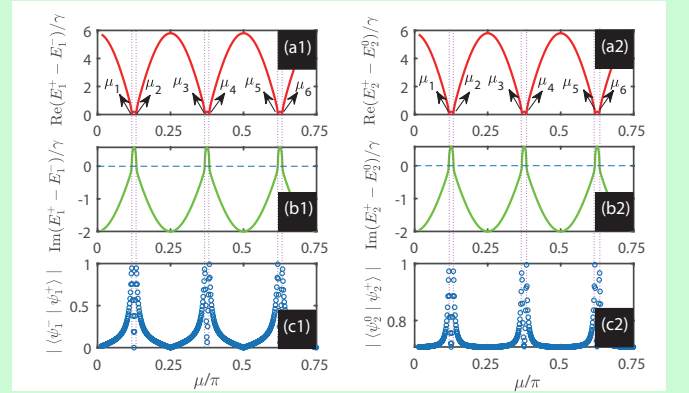}}
   \caption{(a1) and (a2) correspond to the real parts of the frequency splitting ($\text{Re}(E_1^{+}-E_1^{-})$ and $\text{Re}(E_2^{+}-E_2^{0})$)  as functions of $\mu$. (b1) and (b2) denote the imaginary parts of the frequency splitting ($\text{Im}(E_1^{+}-E_1^{-})$ and $\text{Im}(E_2^{+}-E_2^{0})$). (c1) and (c2) show the scalar product between the eigenstates associated with the Hamiltonian~({\ref{hamiltonian5}}) without the driving field as a function of $\mu$. The parameters chosen are $\lambda_1/\gamma=1.5-0.355i$, $\lambda_2/\gamma=1.4-0.645i$, $m=4$, and $U/\gamma=2$.}\label{figure31}
\end{figure}

Different from the degeneracy of eigenenergies, EPs correspond to the situation where both two eigenenergies and their eigenstates coalesce \cite{Konotop88035002,Wiersig033809}. To find EPs of the non-Hermitian system, we plot the real and imaginary parts of the frequency splitting, and the scalar product between the eigenstates associated with the Hamiltonian~({\ref{hamiltonian5}}) without the driving field as functions of $\mu$, as shown in Fig.~\ref{figure31}, which manifests $\hat{H}_{\rm{eff}}$ has two EPs (e.g., $\mu_1$, $\mu_2$ in Fig.~\ref{figure31}) with the energy $E_1^\pm = \omega - U$. In this case, EPs emerge when $E_1^+= E_1^-$, which leads to $\mathcal {E}_{2}$ or $\mathcal {E}_{1}$ equalling zero.
The case of $\mathcal {E}_{1} \ne 0$ and $\mathcal {E}_{2} = 0$ corresponds to solely CW propagation, where the eigenstate is composed of only a single Fock state, i.e., $| {\psi _1^ \pm } \rangle { = } | {1,0} \rangle $, while the case of $\mathcal {E}_{1} = 0$ and $\mathcal {E}_{2} \ne 0$ is associated with the CCW propagation.
From $\mathcal {E}_{2} = 0$ or $\mathcal {E}_{1} = 0$ (resulting in $| {{\lambda _1}} | = | {{\lambda _2}} |$), we obtain
 \begin{equation}
\begin{aligned}
 {\mu }{ = }[n\pi \pm {\arg ({\lambda _1}/{\lambda _2})}]/{(2m)},\, \, \, \, n=\pm 1,\pm 3, \cdots,
\label{exceptionalpointsbeta}
\end{aligned}
 \end{equation}
 where $+$ corresponds to $\mathcal {E}_{1} = 0$ as shown $\mu_2=0.1329\pi$, $\mu_4=0.3829\pi$, and $\mu_6=0.6329\pi$ in Fig.~\ref{figure31}, while $-$ denotes $\mathcal {E}_{2} = 0$ as seen $\mu_1=0.1171\pi$, $\mu_3=0.3671\pi$, and $\mu_5=0.6171\pi$ in Fig.~\ref{figure31}.

\begin{figure}
  \centerline{
  \includegraphics[width=7.2cm, height=7.5cm, clip]{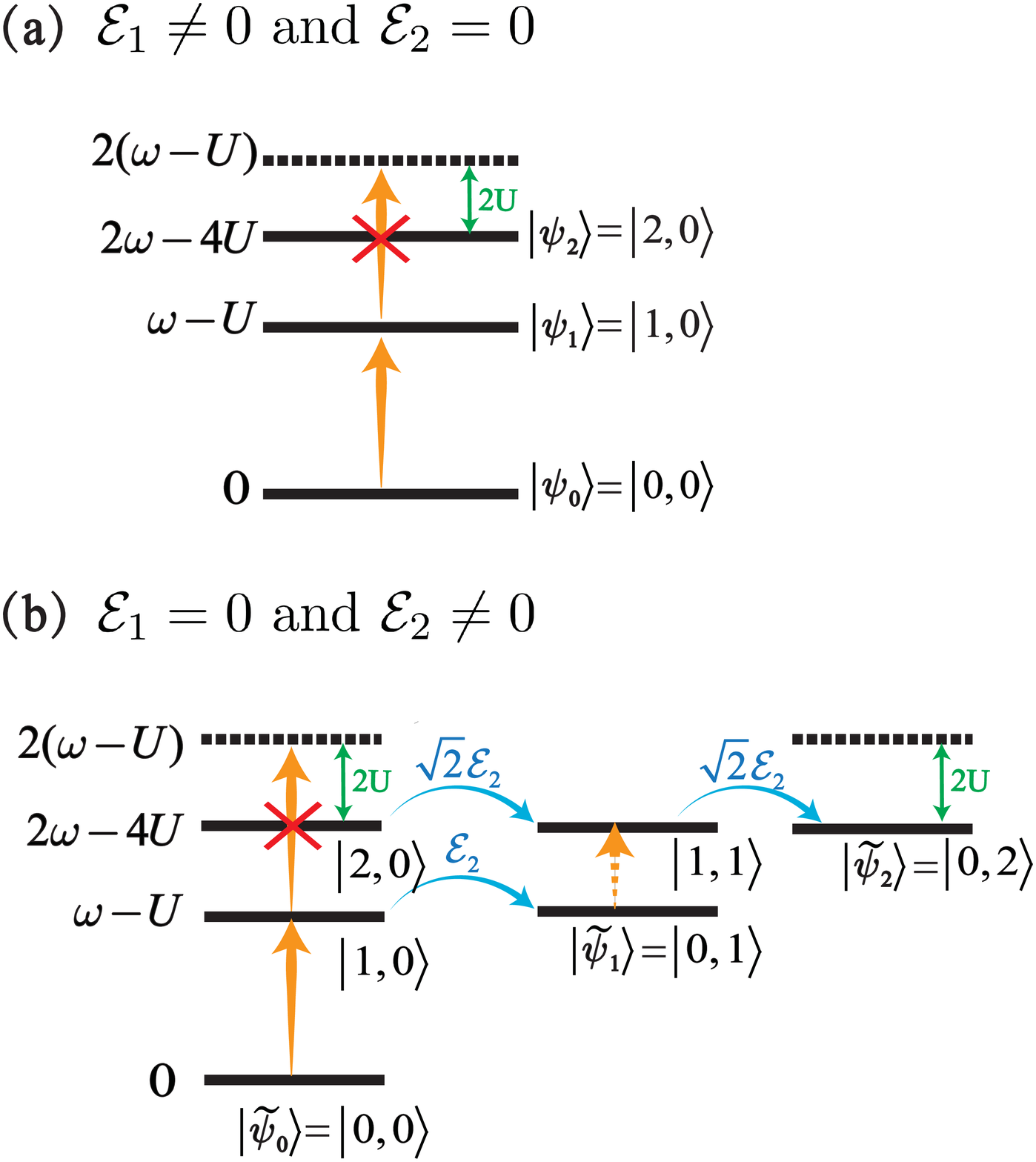}}
   \caption{The energy-level diagram shows the origin of CPB at EPs with the eigenenergy resonance of the single-excitation subspace driven by a laser field (i.e., $\Delta=U$ or $\delta_1=0$), where $| {{\psi _j}} \rangle$ (for $\mathcal {E}_1 \ne 0$ and $\mathcal {E}_2=0$) and $| {\widetilde \psi _j} \rangle$ (for $\mathcal {E}_1=0$ and $\mathcal {E}_2 \ne 0$) denote the eigenstates of Eq.~(\ref{hamiltonian5}) without the driving term. (a) For $\mathcal {E}_1 \ne 0$ and $\mathcal {E}_2=0$ (i.e., the scattering from the CW mode to the CCW mode is forbidden), CPB emerges due to the anharmonic energy levels. (b) For $\mathcal {E}_1=0$ and $\mathcal {E}_2 \ne 0$ (i.e., the scattering from the CCW mode to the CW mode is forbidden), CCW photons are populated in the resonator, which also results in the occurrence of CPB for the CW mode.}\label{figure4}
\end{figure}
Coincidentally, the special $\mu$ in Fig.~\ref{figure1} happens at EPs $\mu_1=0.1171\pi$ and $\mu_2=0.1329\pi$ as shown in Fig.~\ref{figure31}, where the photon statistical properties become extremely interesting. We first discuss the case of $\mathcal {E}_1 \ne 0$ and $\mathcal {E}_2=0$ in Fig.~\ref{figure4}(a).
In this case, the eigenenergy of the system is $E_1=\omega-U$ with the unique eigenstate $| {{\psi _1}} \rangle = | {{1,0}}\rangle$ in the single-excitation subspace. Furthermore, in the two-excitation subspace, the eigenstate $| {{\psi _2}}\rangle = | {{2,0}} \rangle$ composed of only a Fock state has exact energy $2\omega-4U$.
Indeed, the indirect paths $ | {\psi _1} \rangle  \xrightarrow{\mathcal {E}_2} | {0,1} \rangle$, $| {\psi _2} \rangle  \xrightarrow{\sqrt 2 \mathcal {E}_2} | {1,1} \rangle $, and $| {1,1} \rangle  \xrightarrow{\sqrt 2 \mathcal {E}_2} | {0,2} \rangle $ are forbidden due to $\mathcal {E}_2=0$ at EPs, which induces predominantly CW propagating mode. Additionally, the direct path to the two-photon state in CW mode is allowed. Note that the uneven spacing of the energy levels is induced by the nonlinearity, which leads to the strong suppression of the absorption of two photons from the incident laser. Therefore, CPB of the CW mode occurs at the EP $\mu_1=0.1171\pi$.

Moreover, a very different situation appears for $\mathcal {E}_1=0$ and $\mathcal {E}_2 \ne 0$ in Fig.~\ref{figure4}(b), where there is a unique eigenstate $| {{\widetilde{\psi} _1}} \rangle = | {{0,1}} \rangle$ consisting of only a single-photon Fock state, whose energy is exactly $\omega-U$ in the single-excitation subspace. The two-excitation eigenstate $| {{\widetilde{\psi} _2}} \rangle = | {{0,2}} \rangle$  formed by only a Fock state has the exact eigenenergy $2\omega-4U$. The indirect paths $ | {\widetilde{\psi} _1} \rangle  \xrightarrow{\mathcal {E}_1} | {1,0} \rangle$, $| {\widetilde{\psi} _2} \rangle  \xrightarrow{\sqrt 2 \mathcal {E}_1} | {1,1} \rangle $, and $| {1,1} \rangle  \xrightarrow{\sqrt 2 \mathcal {E}_1} | {2,0} \rangle $ are blocked due to $\mathcal {E}_1=0$ at EPs. In other words, only the CW mode couples to the CCW mode, while the CCW mode cannot couple to the CW mode for $\mathcal {E}_1=0$, which suggests CCW propagating mode is predominant.
$| {\widetilde{\psi} _1} \rangle  = | {0,1} \rangle $ means that the CW excitation induced by the driving field is scattered to the CCW mode, and the same is true for the two-excitation subspace. Additionally, the transition from $| {{1,0}} \rangle$ to $| {{2,0}} \rangle$ is forbidden due to the Kerr-type nonlinearity. Therefore, the probability of the second photon in the CW mode is suppressed, which indicates the presence of EPs results in the occurrence of CPB for the CW mode.

Indeed, for $\mathcal {E}_{1} \ne 0$ and $\mathcal {E}_{2} = 0$, the physics is dominated by the photon of CW mode since the eigenstates in the single- and two-excitation subspaces are composed of only a single Fock state $| {n_1,0} \rangle$ of CW mode. For $\mathcal {E}_{1} = 0$ and $\mathcal {E}_{2} \ne 0$, CCW photons are populated in the resonator due to the eigenstates consisting of only a Fock state $| {0,n_2} \rangle$ of CCW mode. The reason for this difference is the nonreciprocal coupling between CW and CCW modes at EPs. Nevertheless, we observe effective CPB for the CW mode in two cases, associated with the suppression of the state $| 2,0 \rangle $. In addition, CPB occurs at EPs if the eigenenergy of the single-excitation subspace is exactly equal to the laser frequency, i.e., $\omega_l= E_1^ \pm =\omega - U$ or $\Delta = \omega- \omega_l=U$ (see below Eq.~(\ref{masterequation})), where the probability of the single-photon state is enhanced. CPB at EPs shows the influence of EPs on the quantum properties of the non-Hermitian system.

\section{comprehensive analysis}
Under the weak driving condition, the state of the system at any time is expanded as
 \begin{equation}
  \begin{aligned}
| {\psi (t)} \rangle  = \sum\limits_{n_1=0,n_2=0}^{N \le 2} {{C_{n_1 n_2}}(t)| {n_1,n_2} \rangle },
\label{expand}
   \end{aligned}
 \end{equation}
where ${C_{n_1 n_2}}(t)$ represents the probability amplitude of the state $| {n_1,n_2} \rangle$, and satisfies ${| {{C_{n_1 n_2}}}|_{N = 2}} \ll {| {{C_{n_1 n_2}}} |_{N = 1}} \ll | {{C_{00}}} | \simeq 1$.
Defining
 \begin{equation}
  \begin{aligned}
   {\delta _1} &= \Delta  - U,\\
   {\delta _2} &= \Delta  - 2U,
   \label{delta1}
   \end{aligned}
 \end{equation}
and substituting Eqs.~({\ref{hamiltonian5}}) and~(\ref{expand}) into Schr\"{o}dinger equation, we obtain the steady-state probability amplitude equations
 \begin{equation}
  \begin{aligned}
     0 &= {\delta _1}{C_{10}} +{\mathcal {E}_1}{C_{01}} + \sqrt 2 F{C_{20}} + F{C_{00}},\\
     0 &= {\delta _1}{C_{01}} + {\mathcal {E}_2}{C_{10}} + F{C_{11}},\\
     0 &= 2{\delta _2}{C_{20}} + \sqrt 2 {\mathcal {E}_1}{C_{11}} + \sqrt 2 F{C_{10}},\\
     0 &= 2{\delta _2}{C_{11}} + \sqrt 2 {\mathcal {E}_1}{C_{02}} + \sqrt 2 {\mathcal {E}_2}{C_{20}} + F{C_{01}},\\
     0 &= 2{\delta _2}{C_{02}} + \sqrt 2 {\mathcal {E}_2}{C_{11}},
  \label{amplitudeeqution}
  \end{aligned}
 \end{equation}
which lead to ${C_{10}}= F{\delta _1}/\eta _1$ and
 \begin{equation}
 \begin{aligned}
   {C_{20}}= \frac{{F^2}(2 \delta_1 \delta_2^2 + \mathcal {E}_1 \mathcal {E}_2 \delta_2 - \mathcal {E}_1 \mathcal {E}_2 \delta_1)}{{2\sqrt 2 {\delta _2}{\eta _1}{\eta _2}}},
  \label{amplitudeexpression}
 \end{aligned}
 \end{equation}
with ignoring higher-order terms in Eq.~(\ref{amplitudeeqution}), where ${\eta _1} = {\mathcal {E}_1}{\mathcal {E}_2} - \delta _1^2$ and ${\eta _2} = {\mathcal {E}_1}{\mathcal {E}_2} - \delta _2^2$. With these results, we obtain
\begin{equation}
\begin{aligned}
 g^{(2)}(0) \simeq \frac{{2{{| {{C_{20}}} |}^2}}}{{{{| {{C_{10}}} |}^4}}}= \frac{| {\eta_1 (2 \delta_1 \delta_2^2 + \mathcal {E}_1 \mathcal {E}_2 \delta_2 - \mathcal {E}_1 \mathcal {E}_2 \delta_1)}| ^2}{4|{ \delta_1^2 \delta_2 \eta_2}| ^2}.
\label{secondordercorrelationfunction}
\end{aligned}
\end{equation}
At EPs ($\mathcal {E}_1 \mathcal {E}_2=0$), when the detuning $\Delta$ is close to $U$ ($\delta_1 \rightarrow 0$), the second-order correlation function tends to zero, i.e.,
 \begin{equation}
\begin{aligned}
  {\left. {{g^{(2)}}(0)} \right|_{{\delta _1} \to 0,{{\cal E}_1}{{\cal E}_2} = 0}} \to 0.
 \label{J1J20delta10}
\end{aligned}
\end{equation}
If the detuning $\Delta$ approaches $U$ ($\delta_1 \rightarrow 0$), but not at EPs ($\mathcal {E}_1 \mathcal {E}_2 \ne 0$), ${g^{(2)}(0)}$ tends to infinite (${g^{(2)}(0)} \rightarrow \infty$), which confirms that the EP is the necessary condition of CPB for this case. Additionally, adjusting the relative angular position $\mu$ to EPs ($\mathcal {E}_1 \mathcal {E}_2=0$), if $\Delta$ is not close to $U$ (equivalently, $\delta_1$ does not tend to zero), ${g^{(2)}(0)}$ can be reduced to $g^{(2)}(0) =\delta_1^2/\delta_2^2$, which may be either greater than one or less than one due to Eq.~(\ref{delta1}).
However, by tuning the parameter $\delta_1$ to $0$ ($\Delta \rightarrow U$) and keeping $\mathcal {E}_1 \mathcal {E}_2=0$, $g^{(2)}(0)$ rapidly decreases to $0$, which indicates the strong antibunching effect. Therefore, the condition of CPB occurring at EPs is given by
 \begin{equation}
  \left\{
  \begin{aligned}
  {\mu _{\rm CPB}} &= [n\pi \pm {\arg ({\lambda _1}/{\lambda _2})}]/{(2m)},\, \, \, \, n=\pm 1,\pm 3,\cdots,\\
  {\Delta _{\rm CPB}} &= U,
\end{aligned}
\right.
 \label{CPBcondition}
\end{equation}
which is consistent with the physical interpretations in Sec. III, where the second equation of Eq.~(\ref{CPBcondition}) can be written as $\delta_1=0$ calculated by Eq.~(\ref{delta1}).

Based on above these, we can explain the relevant CPB phenomena in Fig.~\ref{figure1}. Although the physical mechanisms in both cases ($\mathcal {E}_1=0$ or $\mathcal {E}_2=0$) are slightly different, the presence of EPs results in the occurrence of PB for the eigenenergy resonance of the single-excitation subspace, and the results of the two cases are almost identical, as shown $\mu_1=0.1171\pi$ and $\mu_2=0.1329\pi$ in Fig.~\ref{figure1}.
The physics behind this similarity can be explained from the analytical solution (\ref{secondordercorrelationfunction}), which is symmetric about $\mathcal {E}_1$ and $\mathcal {E}_2$. In other words, $g^{(2)}(0)$ depends on the product of $\mathcal {E}_1$ and $\mathcal {E}_2$ more than their individual. At EPs ($\mathcal {E}_{1} = 0$ or $\mathcal {E}_{2} = 0$), $\mathcal {E}_1 \mathcal {E}_2$ equals zero, which leads to the values of $g^{(2)}(0)$ being almost the same in both cases.

 \begin{figure}[b]
   \centerline{
   \includegraphics[width=8.8cm, height=5.3cm, clip]{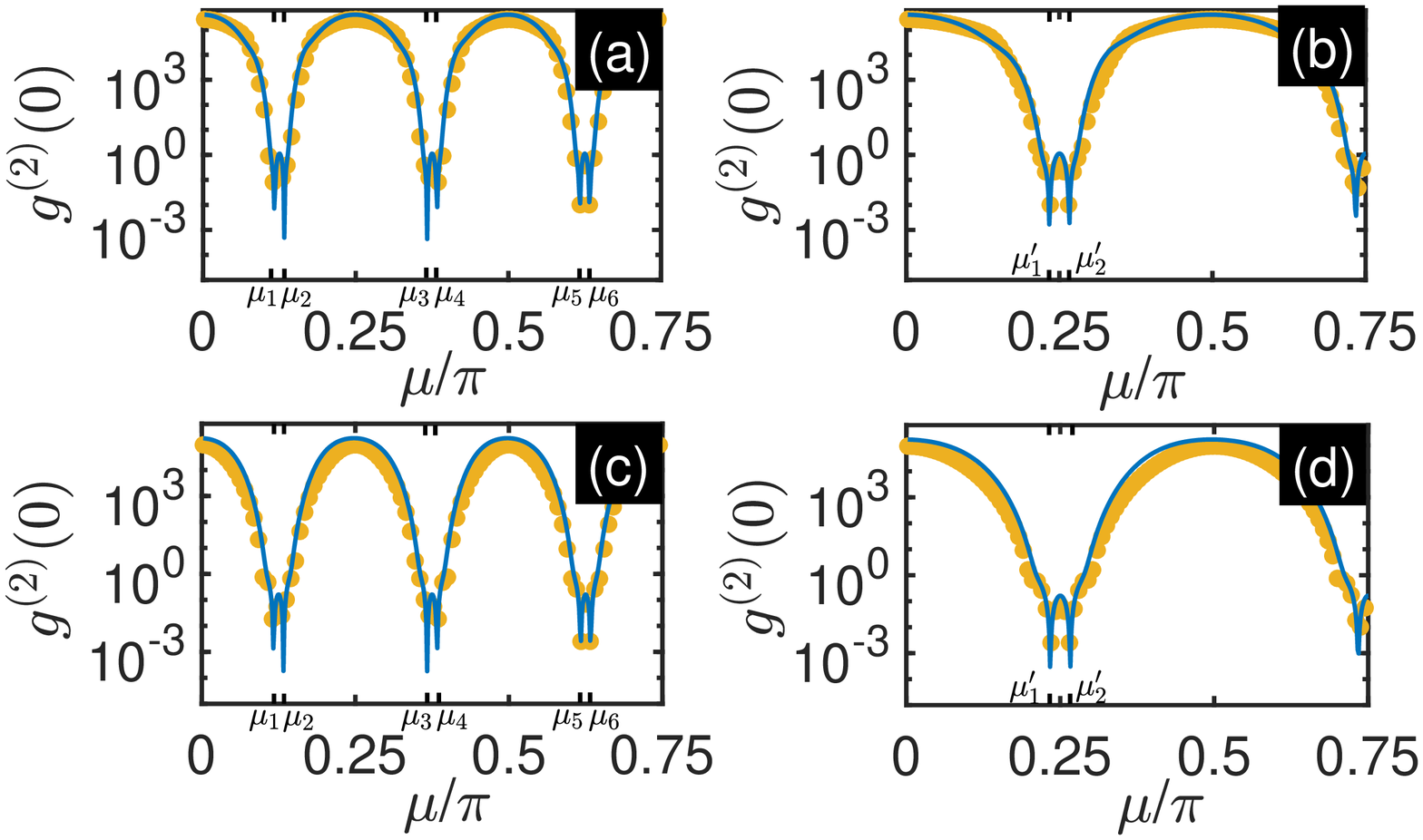}}
   \caption{This figure shows the influences of the different parameters on CPB at EPs by calculating $g^{(2)}(0)$ as a function of $\mu$. The blue-solid lines and yellow dots correspond to the analytical solutions given by Eq.~(\ref{secondordercorrelationfunction}) and numerical simulations in Eq.~(\ref{masterequation}), respectively. The parameters chosen are (a) $\Delta=U=2\gamma$, $m=4$; (b) $\Delta=U=2\gamma$, $m=2$; (c) $\Delta=U=4\gamma$, $m=4$; (d) $\Delta=U=4\gamma$, $m=2$. The other parameters are the same as Fig.~\ref{figure1}(a). $\mu_j \, (j=1,2,\cdots,6)$ corresponding to EPs is the same as Fig.~\ref{figure31}, and $\mu'_1=0.2343\pi$, $\mu'_2=0.2657\pi$. }\label{figure5}
 \end{figure}
Moreover, we plot the analytical and numerical results for $g^{(2)}(0)$ in Fig.~\ref{figure5}, and find that the minimum of $g^{(2)}(0)$ periodically appears with the increase of $\mu$, which implies that CPB periodically exists at EPs.
In Fig.~\ref{figure5}(a) and~(b), when the azimuthal mode number $m$ is twice as small, the period of the line is twice as large, as also reflected in Fig.~\ref{figure5}(c) and~(d), which is because $m$ determines the period of $\mathcal {E}_1$ and $\mathcal {E}_2$ in Eq.~(\ref{nonreciprocal}).
As seen by Fig.~\ref{figure5}(a) and~(c), the optimal angle $\mu$ corresponding to CPB occurring remains unchanged as $U$ varies, which originates from the fact that $U$ does not affect the condition (\ref{CPBcondition}) in case of imposing $\Delta=U$, but it can enhance the performance of PB (also see Fig.~\ref{figure5}(b) and~(d)).
\textbf{The more photon statistical properties of CPB at EPs can be found in Appendix {\ref{SupplementaryCPB}}.}
\begin{figure}
  \centerline{
  \includegraphics[width=7.8cm, height=7.5cm, clip]{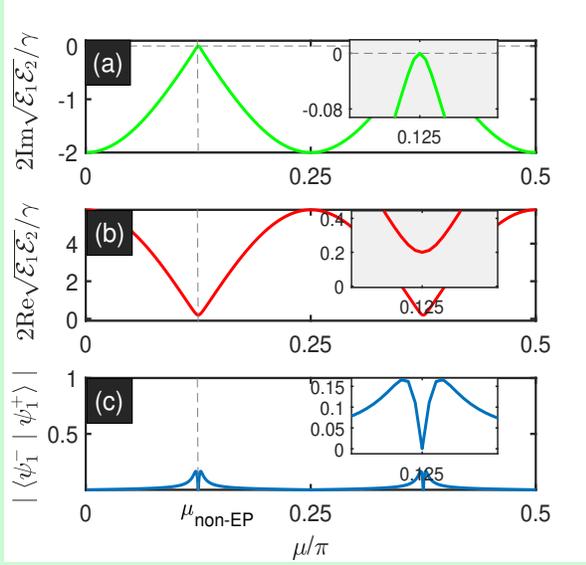}}
   \caption{(a) and (b) correspond to the imaginary and real parts of the complex eigenenergy splitting (${\rm Im} (E_1^ +  - E_1^ - ) = 2\text{Im}\sqrt{\mathcal{E}_1 \mathcal{E}_2}$ and ${\rm Re} (E_1^ +  - E_1^ - ) = 2\text{Re} \sqrt{\mathcal{E}_1 \mathcal{E}_2}$ originated from Eq.~(\ref{E1})) as functions of $\mu$, respectively. (c) shows the scalar product between the eigenstates given by Eq.~(\ref{psi_1}) as a function of $\mu$, where $\mu_{\rm{non-EP}}=0.125\pi$. The parameters chosen are $\lambda_1/\gamma=1.5-0.5i$, $\lambda_2/\gamma=1.4-0.5i$, $m=4$, and $U/\gamma=2$.}\label{except_EP}
\end{figure}

\begin{figure}
  \centerline{
  \includegraphics[width=8.5cm, height=5.8cm, clip]{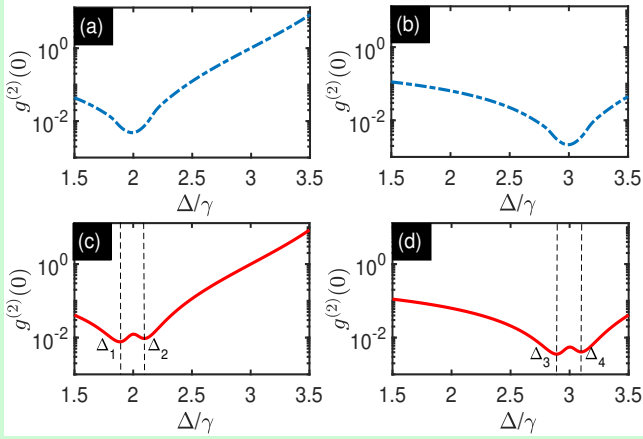}}
   \caption{(a) and (b) show $g^{(2)}(0)$ given by Eq.~(\ref{masterequation}) versus $\Delta$ in order to study CPB at the EP $\mu_1=0.1171\pi$ (see below Eq.~(\ref{exceptionalpointsbeta})), where the parameters chosen are $\lambda_1/\gamma=1.5-0.355i$, $\lambda_2/\gamma=1.4-0.645i$, (a) $U=2\gamma$; (b) $U=3\gamma$. (c) and (d) correspond to $g^{(2)}(0)$ calculated by Eq.~(\ref{masterequation}) as a function of $\Delta$ for investigating CPB at the non-EP $\mu_{\rm{non-EP}}=0.125\pi$ (obtained by Eq.~(\ref{non_EP})), where the parameters chosen are $\lambda_1/\gamma=1.5-0.5i$, $\lambda_2/\gamma=1.4-0.5i$, (c) $U=2\gamma$; (d) $U=3\gamma$. Based on the above parameters, we obtain $\Delta_1=1.9\gamma$, $\Delta_2=2.1\gamma$, $\Delta_3=2.9\gamma$, and $\Delta_4=3.1\gamma$ given by Eq.~(\ref{nonEPCPBcondition}). The other parameters are the same as Fig.~\ref{figure1}(a).}\label{correlation4}
\end{figure}

Before concluding the section, we present a discussion of CPB at non-EPs. Considering that the occurrence of CPB needs to meet the single-excitation eigenenergy resonance, the driven laser frequency $\omega_l$ (real number) is exactly equal to the eigenenergy in Eq.~(\ref{E1}) (i.e., $\omega_l=\omega-U \pm \sqrt{{\mathcal {E}_1}{\mathcal {E}_2}}$ or $\Delta-U \pm \sqrt{{\mathcal {E}_1}{\mathcal {E}_2}} = 0$ see below Eq.~(\ref{masterequation})), which requires the eigenenergies $E_1^{+}$ and $E_1^{-}$ are real. Therefore, the condition of CPB at non-EPs is given by
 \begin{equation}
  \left\{
  \begin{aligned}
  0&=\text{Im} \sqrt{{\mathcal {E}_1}{\mathcal {E}_2}},\\
  \Delta &=U \mp \text{Re} \sqrt{{\mathcal {E}_1}{\mathcal {E}_2}},
\end{aligned}
\right.
 \label{nonEPCPBcondition}
\end{equation}
which can lead to
\begin{equation}
\begin{aligned}
 &\cos 2m\mu  = \\
 &\frac{{\left| {{\mathcal{E}_1}} \right|\left| {{\mathcal{E}_2}} \right| - {{\left| {{\lambda _1}} \right|}^2}({{\cos }^2}{\theta _1} - {{\sin }^2}{\theta _1}) - \left| {{\lambda _2}} \right|({{\cos }^2}{\theta _2} - {{\sin }^2}{\theta _2})}}{{2\left| {{\lambda _1}} \right|\left| {{\lambda _2}} \right|(\cos {\theta _1}\cos {\theta _2} - \sin {\theta _1}\sin {\theta _2})}},
\label{non_EP}
\end{aligned}
\end{equation}
where $\theta _j=\text{arg} \lambda_j$ denotes the argument of the complex number $\lambda_j$. \textbf{In order to consider non-EPs (i.e., $ {\mathcal {E}_1}{\mathcal {E}_2} \ne 0$), we adjust $\lambda_1/\gamma=1.5-0.355i$ and $\lambda_2/\gamma=1.4-0.645i$ (satisfying Eq.~(\ref{exceptionalpointsbeta}) at EPs in Sec.~III) to $\lambda_1/\gamma=1.5-0.5i$ and $\lambda_2/\gamma=1.4-0.5i$, which meet the requirement for non-EPs given by Eq.~(\ref{nonEPCPBcondition}).} Fig.~\ref{except_EP}(a) shows the imaginary part of the complex eigenenergy splitting $2\sqrt{{\mathcal {E}_1}{\mathcal {E}_2}}$ equals zero at $\mu_{\rm{non-EP}}=0.125\pi$, which is consistent with that calculated from Eq.~(\ref{non_EP}). In this case, the real part of $2\sqrt{{\mathcal {E}_1}{\mathcal {E}_2}}$ is not zero (i.e., $2\text{Re} \sqrt{{\mathcal {E}_1}{\mathcal {E}_2}} = 0.2 \gamma$ as shown in the insert of Fig.~\ref{except_EP}(b)) and the scalar product between the eigenstates tends to zero in Fig.~\ref{except_EP}(c), which implies that it is a non-EP.

Fig.~\ref{correlation4}(a) and (b) show $g^{(2)}(0)$ versus the detuning at the EP $\mu_1= 0.1171\pi$ with $\lambda_1/\gamma=1.5-0.355i$ and $\lambda_2/\gamma=1.4-0.645i$, while Fig.~\ref{correlation4}(c) and (d) correspond to $g^{(2)}(0)$ at the non-EP $\mu_{\rm{non-EP}} = 0.125\pi$ with $\lambda_1/\gamma=1.5-0.5i$ and $\lambda_2/\gamma=1.4-0.5i$. If the laser frequency $\omega_l$ is exactly equal to the eigenenergy $E_1^{\pm}$, CPB emerges at the non-EP $\mu_{\rm{non-EP}} = 0.125\pi$ in Fig.~\ref{correlation4}(c) as shown $\Delta_1 = 1.9\gamma$ and $\Delta_2=2.1 \gamma$, which are obtained from $\Delta=U \mp \text{Re} \sqrt{{\mathcal {E}_1}{\mathcal {E}_2}}$ in Eq.~(\ref{nonEPCPBcondition}) with $U=2\gamma$ and $\text{Re} \sqrt{{\mathcal {E}_1}{\mathcal {E}_2}}=0.1\gamma$ given by Fig.~\ref{except_EP}(b). In this case, we show that the eigenstate in the single-excitation subspace splits from one (coalescence) at EPs to two ($| {\psi _1^ - } \rangle$ and $| {\psi _1^ + } \rangle$) at non-EPs.
If $U=3\gamma$, $\text{Im} \sqrt{{\mathcal {E}_1}{\mathcal {E}_2}}$ and $\text{Re} \sqrt{{\mathcal {E}_1}{\mathcal {E}_2}}$ remain unchanged due to $U$ not affecting scattering rates ${\mathcal {E}_{1}}$ and ${\mathcal {E}_{2}}$ given by Eq.~(\ref{nonreciprocal}), which results in the occurrence of CPB at the optimal detunings $\Delta_3=2.9\gamma$ and $\Delta_4=3.1\gamma$ in Fig.~\ref{correlation4}(d).
\begin{figure}
\centering
\includegraphics[height=5.2cm,width=8.8cm]{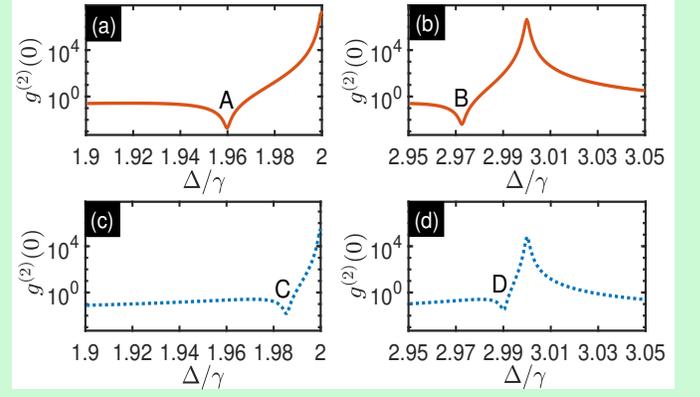}
\caption{\textbf{The figure shows that UPB can occur at non-EPs, where $g^{(2)}(0)$ as a function of $\Delta$ is plotted by solving Eq.~(\ref{masterequation}). Based on Eqs.~({\ref{muUPB}}) and~({\ref{deltaUPB}}), the parameters chosen are $\lambda_1/\gamma=1.5-0.5i$, $U=2\gamma$, $\mu_{\rm{UPB}}=0.1165\pi$, ${\Delta_{\rm{UPB}}}=1.9598\gamma$ for (a); $\lambda_1/\gamma=1.5-0.5i$, $U=3\gamma$, $\mu_{\rm{UPB}}=0.1165\pi$, ${\Delta_{\rm{UPB}}}=2.9726\gamma$ for (b); $\lambda_1/\gamma=1.6-0.5i$, $U=2\gamma$, $\mu_{\rm{UPB}}=0.1132\pi$, ${\Delta_{\rm{UPB}}}=1.9855\gamma$ for (c); $\lambda_1/\gamma=1.6-0.5i$, $U=3\gamma$, $\mu_{\rm{UPB}}=0.1132\pi$, ${\Delta_{\rm{UPB}}}=2.9903\gamma$ for (d). The other parameters chosen are $\lambda_2/\gamma=1.4-i$, $F=0.1\gamma$, and $m=4$.}}
\label{UPB}
\end{figure}

\begin{figure*}
\centerline{
\includegraphics[width=16.5cm, height=4cm, clip]{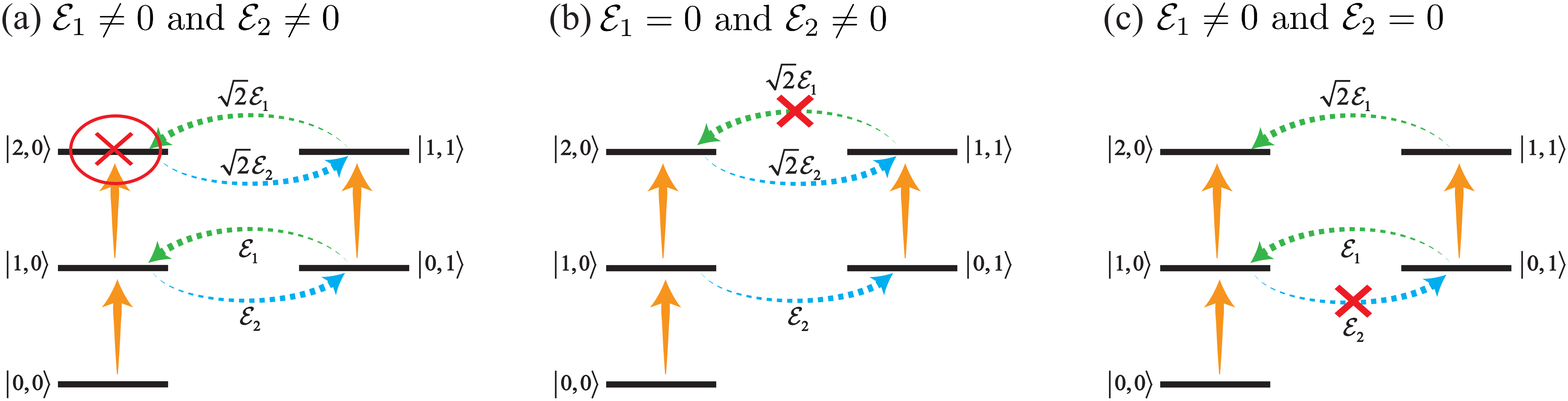}}
\caption{\textbf{(a) The transition paths at non-EPs (${\mathcal {E}_1} \ne 0$ and ${\mathcal {E}_2} \ne 0$) lead to the destructive quantum interference responsible for UPB. (b) and (c) correspond to the transition paths at EPs for $\mathcal {E}_1 = 0$ and $\mathcal {E}_2 = 0$, respectively, which reveal that the destructive quantum interference cannot occur due to the two different quantum pathways to the two-photon state being not formed.}}\label{transition}
\end{figure*}
\section{discussions on UPB}
\textbf{In order to give a complete description of the non-Hermitian system, we discuss UPB at non-EPs and EPs.}

\textbf{(1) UPB can exist at non-EPs (${\mathcal {E}_1}{\mathcal {E}_2} \ne 0$), where the optimal condition can be derived by setting $C_{20}=0$ in Eq.~(\ref{amplitudeexpression}). Taking the imaginary part of Eq.~(\ref{amplitudeexpression}) as zero, we have}
\begin{equation}
\begin{aligned}
 {\mathop{\rm Im}\nolimits} ({\mathcal {E}_1}{\mathcal {E}_2}) = 0,
\label{imaginary}
\end{aligned}
\end{equation}
\textbf{while the real part of Eq.~(\ref{amplitudeexpression}) equalling zero leads to}
\begin{equation}
\begin{aligned}
 {\mathop{\rm Re}\nolimits} ({\mathcal {E}_1}{\mathcal {E}_2}) = \frac{{2{\delta _1}\delta _2^2}}{{{\delta _1} - {\delta _2}}},
\label{real}
\end{aligned}
\end{equation}
\textbf{which induces $\delta_1 \ne 0$ (or $\Delta \ne U$) due to ${\mathop{\rm Re}\nolimits} ({\mathcal {E}_1}{\mathcal {E}_2}) \ne 0$. Substituting Eq.~(\ref{nonreciprocal}) into Eq.~(\ref{imaginary}), we have}
\begin{equation}
  \begin{aligned}
   \cos (2m\mu_{\rm{UPB}}) =  - \frac{{{{| {{\lambda _1}} |}^2}\sin {\theta _1}\cos {\theta _1} + {{| {{\lambda _2}} |}^2}\sin {\theta _2}\cos {\theta _2}}}{{| {{\lambda _1}}| | {{\lambda _2}} |\sin ({\theta _1} + {\theta _2})}}.
    \label{muUPB}
  \end{aligned}
 \end{equation}
\textbf{With Eqs.~(\ref{muUPB}) and~(\ref{nonreciprocal}), Eq.~(\ref{real}) gives}
 \begin{equation}
  \begin{aligned}
   {\Delta_{\rm{UPB}}} = \frac{{{{\left( { - 2} \right)}^{4/3}}{U^2} + {{\left( { - 2} \right)}^{2/3}}{M^2} + 10UM}}{{6M}},
    \label{deltaUPB}
  \end{aligned}
 \end{equation}
\textbf{where}
\begin{equation}
  \begin{aligned}
   M = {[  3( {  \sqrt {1344{U^6} + 660{U^3}q + 81{q^2}} } - 9q )} { - 110{U^3} ]^{1/3}}, \nonumber
  \end{aligned}
 \end{equation}
\textbf{with $q =  - 4{U^3} - U{\mathop{\rm Re}\nolimits} ({{\mathcal {E}_1}{\mathcal {E}_2}})$. We notice that when the optimal conditions in Eqs.~(\ref{muUPB}) and~(\ref{deltaUPB}) simultaneously are satisfied, the strong antibunching can be obtained, otherwise the system is not in the strong antibunching regime.}

\textbf{In Fig.~\ref{UPB}(a), taking $\lambda_1/\gamma=1.5-0.5i$, $\lambda_2/\gamma=1.4-i$, $m=4$, and $U=2\gamma$, UPB occurs at $\mu_{\rm{UPB}}=0.1165\pi$ and ${\Delta_{\rm{UPB}}}=1.9598\gamma$ given by Eqs.~(\ref{muUPB}) and~(\ref{deltaUPB}), which explain the point A, while $\mu_{\rm{UPB}}=0.1165\pi$ and ${\Delta_{\rm{UPB}}}=2.9726\gamma$ at $U=3 \gamma$ are given as the point B in Fig.~\ref{UPB}(b). Moreover, the point C in Fig.~\ref{UPB}(c) corresponds to the optimal conditions $\mu_{\rm{UPB}}=0.1132\pi$ and ${\Delta_{\rm{UPB}}}=1.9855\gamma$ for $U=2\gamma$ calculated by Eqs.~({\ref{muUPB}}) and~({\ref{deltaUPB}}) when we change $\lambda_1=(1.5-0.5i)\gamma$ to $\lambda_1=(1.6-0.5i)\gamma$. The point D in Fig.~\ref{UPB}(d) is evaluated by $\mu_{\rm{UPB}}=0.1132\pi$ and ${\Delta_{\rm{UPB}}}=2.9903\gamma$ at $U=3\gamma$. The energy levels and transition paths are shown in Fig.~\ref{transition}(a), where the non-zero nonreciprocal scattering rates ${\mathcal {E}_1}$ and ${\mathcal {E}_2}$ lead to the occurrence of the destructive quantum interference between two different excitation paths.}

\textbf{(2) UPB does not exist at EPs, which is discussed as follows.}

\textbf{(i) The destructive quantum interference between two (or more) different excitation paths occurs for the existence of UPB. With Eq.~(\ref{amplitudeeqution}), $\mathcal {E}_1 = 0$ leads to}
\begin{equation}
  \begin{aligned}
     0 &= {\delta _1}{C_{10}} + F{C_{00}},\\
     0 &= {\delta _1}{C_{01}} + {\mathcal {E}_2}{C_{10}},\\
     0 &= 2{\delta _2}{C_{20}} + \sqrt 2 F{C_{10}},\\
     0 &= 2{\delta _2}{C_{11}} + \sqrt 2 {\mathcal {E}_2}{C_{20}} + F{C_{01}},\\
     0 &= 2{\delta _2}{C_{02}} + \sqrt 2 {\mathcal {E}_2}{C_{11}},\label{E10}
  \end{aligned}
 \end{equation}
\textbf{which corresponds to Fig.~\ref{transition}(b). We note that there is only one path for the system to reach the two-photon state $|2,0\rangle$ given by Eq.~(\ref{E10}) of CW mode, i.e., the direct transition $|1,0\rangle \rightarrow |2,0\rangle$.}

\textbf{For $\mathcal {E}_2=0$, we have}
\begin{equation}
  \begin{aligned}
     0 &= {\delta _1}{C_{10}} +{\mathcal {E}_1}{C_{01}} + F{C_{00}},\\
     0 &= {\delta _1}{C_{01}},\\
     0 &= 2{\delta _2}{C_{20}} + \sqrt 2 {\mathcal {E}_1}{C_{11}} + \sqrt 2 F{C_{10}},\\
     0 &= 2{\delta _2}{C_{11}} + \sqrt 2 {\mathcal {E}_1}{C_{02}}+ F{C_{01}},\\
     0 &= 2{\delta _2}{C_{02}},\label{E20}
  \end{aligned}
 \end{equation}
\textbf{where the steady-state solution $C_{01}=C_{02}=C_{11}=0$ can be obtained. Therefore, the transition to the two-photon state $|2,0\rangle$ only contains $|1,0\rangle \rightarrow |2,0\rangle$ in Fig.~\ref{transition}(c). The destructive quantum interference cannot occur due to Eqs.~(\ref{E10}) and (\ref{E20}), which results in UPB not existing.}

\textbf{(ii) UPB requires $C_{20} =0$. At EPs (${\mathcal {E}_1} {\mathcal {E}_2} = 0$ or $\text{Im} ({\mathcal {E}_1} {\mathcal {E}_2})= \text{Re} ({\mathcal {E}_1} {\mathcal {E}_2})=0$), the optimal conditions ($\text{Im}(C_{20})=0$ and $\text{Re}(C_{20})=0$) given by Eqs.~(\ref{imaginary}) and (\ref{real}) for UPB existing lead to $\delta_1=0$, which is exactly consistent with Eq.~(\ref{CPBcondition}) (i.e., the condition of CPB occurring).}

\textbf{In summary, we point out that UPB exists only if the following two conditions are simultaneously satisfied: (I) The destructive quantum interference between two (or more) different excitation paths occurs; (II) The two-photon probability amplitude $C_{20}$ equals zero. To be specific, the first condition (I) is not met in the case labeled by (i), which leads to the fact that UPB does not exist at EPs.}

\section{conclusions and discussions}
In conclusion, we have found that CPB emerges at EPs of the non-Hermitian optomechanical system coupled with the driven WGM microresonator under the weak optomechanical coupling approximation for the eigenenergy resonance of the single-excitation subspace driven by a laser field.
We also discuss the origin of CPB at EPs.
For $\mathcal {E}_1 \ne 0$ and $\mathcal {E}_2 = 0$, with the direct path to the two-photon state in CW mode allowed, CPB emerges due to the anharmonic energy level, where the eigenstates are composed of only a Fock state $| {n_1,0} \rangle$.
For $\mathcal {E}_1 = 0$ and $\mathcal {E}_2 \ne 0$, the coalescence of eigenstates consisting of only a Fock state $| {0,n_2} \rangle$ causes the system unable to absorb the second photon in the CW mode, which also results in the occurrence of CPB.

Moreover, we study CPB at non-EPs, which occurs for two optimal detunings because of the eigenstate splitting from one (coalescence) at EPs to two at non-EPs in the single-excitation subspace.
\textbf{We also show that UPB does not exist at EPs since the destructive quantum interference cannot occur due to the two different quantum pathways to the two-photon state being not formed, while UPB can occur at non-EPs.}
In a broader view, our results may have important applications in generating single-photon sources for the non-Hermitian optomechanical system, which aims to improve the performance of quantum sensors \cite{Park462,Zhang180501,Wiersig1} and quantum unidirectional devices \cite{Peng1136845,Yin175,Shen125203}.

\section*{ACKNOWLEDGMENTS}
This work was supported by National Natural Science Foundation of China under Grants No. 12274064, and Natural Science Foundation of Jilin Province (subject arrangement project) under Grant No. 20210101406JC.

\section*{}
\appendix
\section{\label{experimental}Discussions on the experimental implementation}
In this section, we present a discussion on the experimental feasibility of observing the prediction for an optomechanical system coupled with the driven WGM microresonator. For the model under study, we mainly focus on the following point: the scatterer-induced nonreciprocal coupling of the CW and CCW traveling lights.

The non-Hermitian optical coupling of the CW and CCW traveling lights can be induced by the nanoparticles, where the isolated microdisk cavity is perturbed by two particles in Refs.\cite{Wiersig063828,Wiersig203901}. To gain more insights into Eq.~(\ref{nonreciprocal}), we briefly review the two-mode approximation model, whose central assumption is that the small perturbation induced by the nanoparticles couples only the modes within a given degenerate mode pair of the isolated microdisk.
The key idea is to model the dynamics in the slowly-varying envelope approximation in the time domain with a Schr\"{o}dinger-type equation. Considering that the WGM microcavity is an open system, the effective Hamiltonian in the traveling-wave basis (CW, CCW) is given by the $2\times2$ non-Hermitian matrix
\begin{eqnarray}
\hat{H} = \left( {\begin{array}{*{20}{c}}
\mathcal{C} &{{\mathcal {E}_2}}\\
{{\mathcal {E}_1}}&\mathcal{C}
\end{array}} \right),
\label{twomodeapproximationmodel}
\end{eqnarray}
where the real and imaginary parts of the diagonal element $\mathcal{C}$ correspond to the frequency of the system and the decay rate of the resonant traveling waves, respectively. The complex-valued off-diagonal element $\mathcal {E}_1$ ($\mathcal {E}_2$) is the backscattering coefficient, which describes the scattering from the CCW (CW) to the CW (CCW) traveling wave. In general, it is possible that the backscattering between CW and CCW traveling waves is asymmetric, i.e., $| {{\mathcal {E}_1}} | \ne | {{\mathcal {E}_2}} |$.
For the particular case of the WGM microresonator perturbed through two scatterers, ignoring the frequency shifts for negative-parity modes, the matrix elements of $\hat{H}$ are determined by
\begin{equation}
\begin{aligned}
\begin{array}{l}
\mathcal{C}  = {\omega _0} - i{\gamma } + \sum\limits_{j = 1}^2 {{\lambda _j}},\\
{\mathcal {E}_1} = \sum\limits_{j = 1}^2 {{\lambda _j}} {e^{i2m{\mu _{S_j}}}},\\
{\mathcal {E}_2} = \sum\limits_{j = 1}^2 {{\lambda _j}} {e^{ -i 2m{\mu _{S_j}}}},
\end{array}
\label{matrixelements}
\end{aligned}
\end{equation}
where $\gamma$ denotes the decay rate. $m$ is the azimuthal mode number. $\mu_{S_j}$ is the angular position of scatterer $S_j$. $\lambda_j$ is the complex frequency splitting induced by scatterer $S_j$ alone. In this case, we take the position of one of the nanoparticles as the reference position. To be specific, we take nanoparticle $S_1$ (see Fig.~\ref{device}) as the first particle in this model and set its angular position to $\mu_{S_1} = 0$. Subsequently, the angular position of the second particle $S_2$ is $\mu_{S_2} = \mu$, where $\mu$ denotes the relative angular position of the two nanoparticles. Therefore, the asymmetric backscattering coefficients of CW and CCW traveling waves induced by the nanoparticles are reduced to
\begin{equation}
\begin{aligned}
\begin{array}{l}
 {\lambda _1} + {\lambda _2}{e^{ \pm i2m\mu }},
\end{array}
\label{J_12}
\end{aligned}
\end{equation}
which are consistent with Eq.~(\ref{nonreciprocal}). The backscattering coefficients can be adjusted by tuning the relative angle $\mu$, which modifies the photon statistical properties of the system.
It is worth noting that ${\lambda _j}$ can be calculated for the single-particle-microdisk system either fully numerically (using, e.g., the finite element method \cite{Braess}, the boundary element method \cite{Wiersig553}) or analytically using the Green's function approach \cite{Dettmann063813}.

\section{\label{DerivationHamiltonian} Derivation of system Hamiltonian}
  The Hamiltonian of the whole system is given by
  \begin{equation}
   \begin{aligned}
   {\hat{H}_{1}} &= {\hat{H}_0} + {\hat{H}_m} + {\hat{H}_{{\mathop{\rm int}} }} + {\hat{H}_{\rm{dr}}},\\
   {\hat{H}_0} &= \omega_1 {\hat{a}_1^\dag} {\hat{a}_1} + \omega_2 {\hat{a}_2^\dag} {\hat{a}_2} + {\mathcal {E}_1}{\hat{a}_1^\dag} {\hat{a}_2} + {\mathcal {E}_2}{\hat{a}_2^\dag} {\hat{a}_1},\\
   {\hat{H}_m} &= \frac{{{\hat{p}^2}}}{{2{m_{\rm{eff}}}}} + \frac{1}{2}{m_{\rm{eff}}}\omega _m^2{\hat{x}^2},\\
   {\hat{H}_{{\mathop{\rm int}} }} &=  - G\hat{x}({\hat{a}_1^\dag} {\hat{a}_1} + {\hat{a}_2^\dag} {\hat{a}_2}),\\
   {\hat{H}_{\rm{dr}}} &= F{e^{ - i{\omega _l}t}}{\hat{a}_1^\dag}  + {F }{e^{i{\omega _l}t}}{\hat{a}_1},
   \label{hamiltonianA1}
   \end{aligned}
  \end{equation}
  where $G = \omega_0 / R$ denotes the cavity optomechanics coupling coefficient \cite{Grudinin083901}. Making a canonical transformation to the annihilation operator $\hat{b}$ and creation operator $\hat{b}^\dag$ as $\hat{x} = \sqrt {1/(2{m_{\rm{eff}}}{\omega _m})} ({\hat{b}^\dag } + \hat{b})$ and $\hat{p} = i\sqrt {{m_{\rm{eff}}}{\omega _m}/2} ({\hat{b}^\dag } - \hat{b})$, Eq.~(\ref{hamiltonianA1}) is reduced to
 \begin{equation}
  \begin{aligned}
  {\hat{H}_{2}} = &{\hat{H}_{\rm{opt}}} + {\hat{H}_{\rm{dr}}},\\
  {\hat{H}_{\rm{opt}}} = &\omega_1 {\hat{a}_1^\dag} {\hat{a}_1} + \omega_2 {\hat{a}_2^\dag} {\hat{a}_2} + {\omega _{m}}{\hat{b}^\dag }\hat{b} + {\mathcal {E}_1}{\hat{a}_1^\dag} {\hat{a}_2} + {\mathcal {E}_2}{\hat{a}_2^\dag} {\hat{a}_1}\\
    &- g({\hat{b}^\dag } + \hat{b})(\hat{a}_1^\dag {\hat{a}_1} + \hat{a}_2^\dag {\hat{a}_2}), \\
  {\hat{H}_{\rm{dr}}} = &F{e^{ - i{\omega _l}t}}{\hat{a}_1^\dag}  + {F }{e^{i{\omega _l}t}}{\hat{a}_1},
   \label{Kerrhamtonian}
  \end{aligned}
 \end{equation}
 where $\omega_j$ is the resonance frequency of the $j$th cavity, and the cavity optomechanics coupling coefficient can be changed as $ g = G /\sqrt{2 m_{\rm{eff}} \omega_{m}}$. By performing a rotating transformation defined by ${{\hat{V}}_1} = \exp [ - i{\omega _l}t(\hat{a}_1^\dag {\hat{a}_1} + \hat{a}_2^\dag {\hat{a}_2})]$, Eq.~(\ref{Kerrhamtonian}) becomes
 \begin{equation}
  \begin{aligned}
   {\hat{H}_{T}} &=  {\hat{V}_1}^\dag {\hat{H}_2}\hat{V}_1 - i {\hat{V}_1}^\dag  \frac{{d{\hat{V}_1}}}{{dt}} \\
   &=\Delta_1 {\hat{a}_1^\dag} {\hat{a}_1} + \Delta_2 {\hat{a}_2^\dag} {\hat{a}_2} + {\omega _{m}}{\hat{b}^\dag }\hat{b} + {\mathcal {E}_1}{\hat{a}_1^\dag} {\hat{a}_2} + {\mathcal {E}_2}{\hat{a}_2^\dag}   {\hat{a}_1}\\
   & \quad - g({\hat{b}^\dag } + \hat{b})(\hat{a}_1^\dag {\hat{a}_1} + \hat{a}_2^\dag {\hat{a}_2}) + F{\hat{a}_1^\dag}  + {F }{\hat{a}_1},\label{hamil}
  \end{aligned}
 \end{equation}
 \textbf{which corresponds to Eq.~(\ref{hamiltonian3}).}
 \textbf{In order to decouple the mechanical resonator from the total Hamiltonian, the Hamiltonian~{(\ref{hamil})} with the time-independent unitary transformation by ${\hat{V}_2} = \exp [{g}/{\omega _m}(\hat{a}_1^\dag {\hat{a}_1} + \hat{a}_2^\dag {\hat{a}_2})({\hat{b}^\dag } - \hat{b})]$ leads to}
 \begin{equation}
  \begin{aligned}
   {\hat{H}_{3}} &= {\hat{V}_2}^\dag {\hat{H}_{T}}\hat{V}_2\\
   &= \Delta_1 {\hat{V}_2}^\dag {\hat{a}_1^\dag} {\hat{a}_1} \hat{V}_2 + \Delta_2 {\hat{V}_2}^\dag {\hat{a}_2^\dag} {\hat{a}_2} \hat{V}_2 + {\omega _{m}} {\hat{V}_2}^\dag {\hat{b}^\dag }\hat{b} \hat{V}_2 \\
   &\quad + {\mathcal {E}_1}{\hat{V}_2}^\dag {\hat{a}_1^\dag} {\hat{a}_2} \hat{V}_2 + {\mathcal {E}_2} {\hat{V}_2}^\dag {\hat{a}_2^\dag} {\hat{a}_1} \hat{V}_2 \\
   &\quad - g {\hat{V}_2}^\dag ({\hat{b}^\dag } + \hat{b}) (\hat{a}_1^\dag {\hat{a}_1} +  \hat{a}_2^\dag {\hat{a}_2})\hat{V}_2 \\
   &\quad + F {\hat{V}_2}^\dag {\hat{a}_1^\dag}\hat{V}_2  + {F } {\hat{V}_2}^\dag {\hat{a}_1}\hat{V}_2. \\.\label{h3B}
   \end{aligned}
  \end{equation}
 \textbf{With ${e^{\alpha \hat{A}}}\hat{B}{e^{ - \alpha \hat{A}}} = \hat{B} + \alpha [\hat{A},\hat{B}] + \frac{{{\alpha ^2}}}{{2!}}[\hat{A},[\hat{A},\hat{B}]] + \cdots$ \cite{Sakurai2017}, we have the following identities}
 \begin{equation}
  \begin{aligned}
   {\hat{V}_2}^\dag {\hat{a}_1}^\dag {\hat{V}_2} &= {\hat{a}_1}^\dag {e^{ - \frac{g}{{{\omega _m}}}({\hat{b}^\dag } - \hat{b})}},\\
   {\hat{V}_2}^\dag {\hat{a}_2}^\dag {\hat{V}_2} &= {\hat{a}_2}^\dag {e^{ - \frac{g}{{{\omega _m}}}({\hat{b}^\dag } - \hat{b})}},\\
   {\hat{V}_2}^\dag {\hat{a}_1} {\hat{V}_2} &= {\hat{a}_1} {e^{  \frac{g}{{{\omega _m}}}({\hat{b}^\dag } - \hat{b})}},\\
   {\hat{V}_2}^\dag {\hat{a}_2} {\hat{V}_2} &= {\hat{a}_2} {e^{  \frac{g}{{{\omega _m}}}({\hat{b}^\dag } - \hat{b})}},\\
   {\hat{V}_2}^\dag \hat{b}^\dag {\hat{V}_2}&= {\hat{b}^\dag } + \frac{g}{{{\omega _m}}}({\hat{a}_1}^\dag {\hat{a}_1} + {\hat{a}_2}^\dag {\hat{a}_2}),\\
   {\hat{V}_2}^\dag \hat{b} {\hat{V}_2}&= {\hat{b}} + \frac{g}{{{\omega _m}}}({\hat{a}_1}^\dag {\hat{a}_1} + {\hat{a}_2}^\dag {\hat{a}_2}),
  \label{guocheng}
   \end{aligned}
  \end{equation}
\textbf{and then obtain ${\hat{H}_{3}}$ from Eq.~(\ref{h3B}) as follows}
   \begin{equation}
  \begin{aligned}
   {\hat{H}_{3}} = &\Delta_1 {\hat{a}_1^\dag} {\hat{a}_1} + \Delta_2 {\hat{a}_2^\dag} {\hat{a}_2} + {\omega _{m}}{\hat{b}^\dag }\hat{b} + {\mathcal {E}_1}{\hat{a}_1^\dag} {\hat{a}_2} + {\mathcal {E}_2}{\hat{a}_2^\dag} {\hat{a}_1}\\
   & -{g^2}/{\omega _{m}}[(\hat{a}_1^\dag {\hat{a}_1})^2 + (\hat{a}_2^\dag {\hat{a}_2})^2 + 2\hat{a}_1^\dag {\hat{a}_1}\hat{a}_2^\dag {\hat{a}_2}] \\
   & + F{e^{g/{\omega _{m}}(\hat{b}-\hat{b}^\dag)}}{\hat{a}_1^\dag}  + {F }{e^{-g/{\omega _{m}}(\hat{b}-\hat{b}^\dag)}}{\hat{a}_1}.\label{Kerrhamtonian1}
   \end{aligned}
  \end{equation}
 Under the weak optomechanical coupling approximation ($g/\omega_m \ll 1$), Eq.~(\ref{hamiltonian4}) can be obtained by approximating $e^{g/{\omega _{m}}(\hat{b}-\hat{b}^\dag)}$ to $1$ in Eq.~(\ref{Kerrhamtonian1}).
\begin{figure}
  \centerline{
  \includegraphics[width=7.9cm, height=7.3cm, clip]{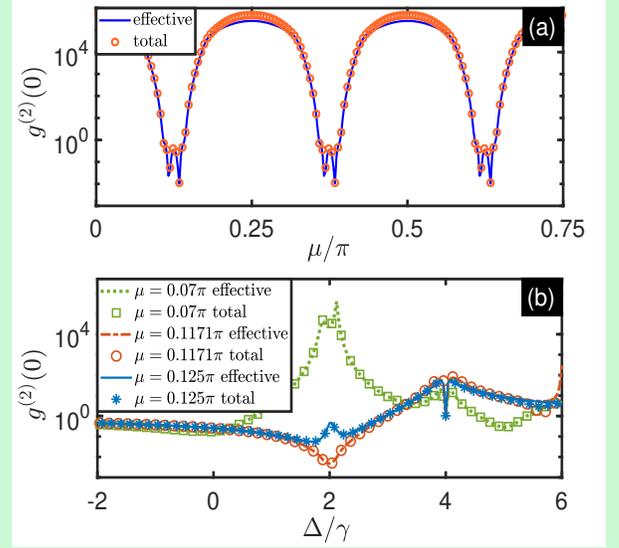}}
  \caption{(a) shows $g^{(2)}(0)$ as a function of $\mu$ for weak optomechanical coupling at fixed detuning $\Delta=U$.  The blue line and red circles correspond to the numerical result based on the effective master equation ({\ref{masterequation}}) and total master equation ({\ref{master_equation_original}}), respectively. (b) corresponds to $g^{(2)}(0)$ as a function of $\Delta$ with various angles $\mu$ for both effective master equation ({\ref{masterequation}}) (different styles of curves) and total master equation ({\ref{master_equation_original}}) (different styles of data point symbols) under the weak optomechanical coupling approximation.  The parameters chosen are $\omega_m=30\gamma$ and $g=7.746\gamma$. The other parameters are the same as Fig.~\ref{figure1}(a).}\label{figure10}
  \end{figure}
\section{\label{validity} the validity of approximate hamiltonian}

In this model, the crucial factor in investigating the photon statistical properties is whether the effective Hamiltonian ({\ref{hamiltonian5}}) is equivalent to the total Hamiltonian ({\ref{hamiltonian3}}) in the case of weak optomechanical coupling. To check the validity of effective Hamiltonian $\hat{H}_{\rm{eff}}$ in Eq.~(\ref{hamiltonian5}), we give the quantum master equation for the optomechanical system
\begin{equation}
\begin{aligned}
 \frac{{d{\rho}_{\rm{opt}} }}{{dt}} =  &- i[ {{\widetilde{H}_ + },{\rho}_{\rm{opt}} } ] - i\{ {{{\widetilde{H}}_ - },{\rho_{\rm{opt}}} } \} + 2i{\rm Tr}({\rho}_{\rm{opt}} {{\widetilde{H}}_ - }){\rho}_{\rm{opt}}\\
  & + \sum\limits_j {({{{\gamma _j}} \mathord{\left/  {\vphantom {{{\gamma _j}} 2}} \right. \kern-\nulldelimiterspace} 2})\mathcal{D}({\rho}_{\rm{opt}},{\hat{a}_j})},
\label{master_equation_original}
\end{aligned}
\end{equation}
where ${\widetilde{H}}_{+}=({\hat{H}_{T}}+\hat{H}_{T}^\dag)/2$ and ${\widetilde{H}}_{-}=({\hat{H}_{T}}-\hat{H}_{T}^\dag)/2$ denote the Hermitian and anti-Hermitian parts of the total Hamiltonian $\hat{H}_T$ given by Eq.~(\ref{hamiltonian3}), respectively.

\begin{figure}
  \centerline{
  \includegraphics[width=8.5cm, height=4.3cm, clip]{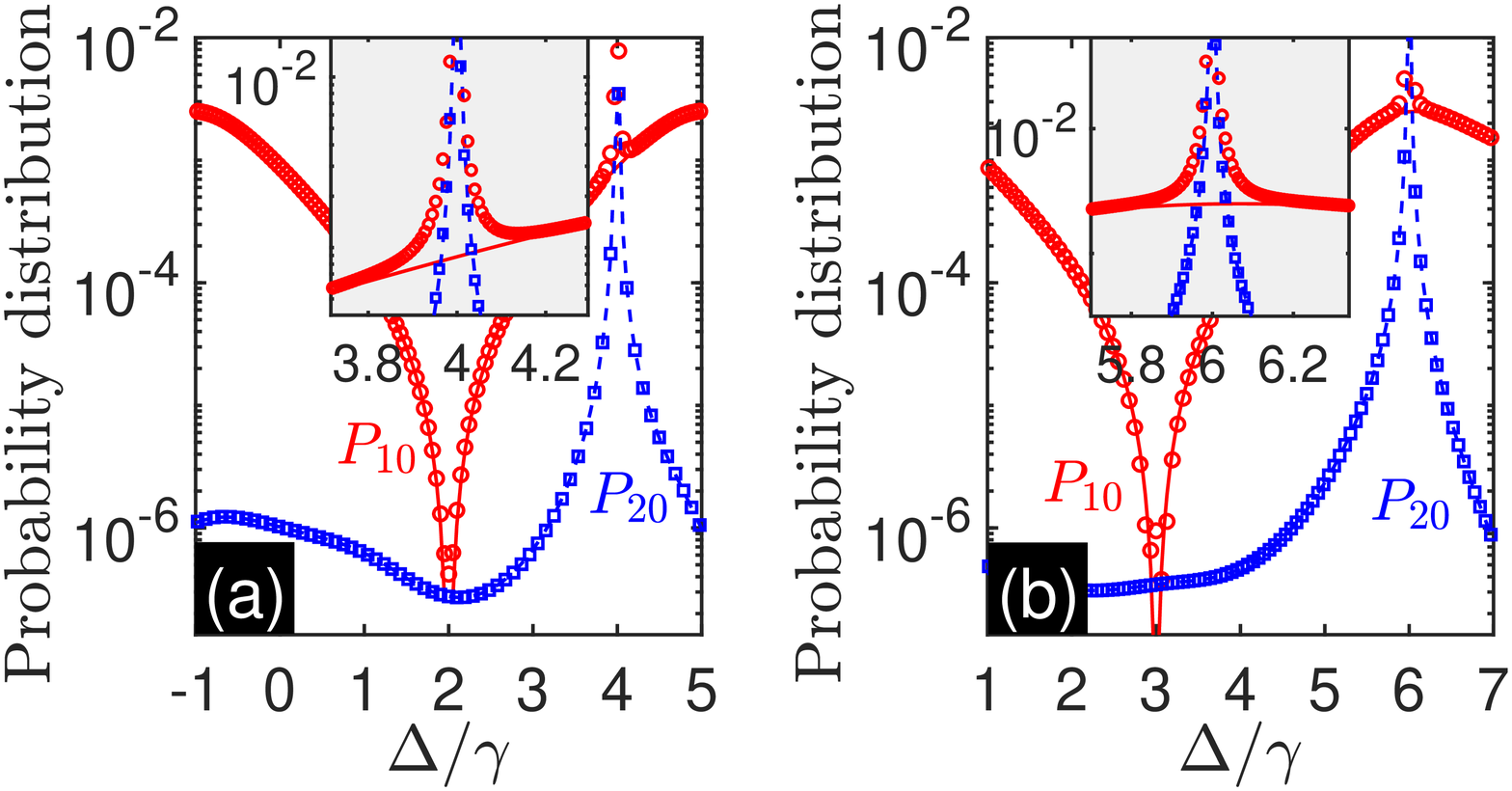}}
  \caption{The photon probability distribution varies with $\Delta$ at $\mu=0$. The solid (dashed) lines and circles (squares) correspond to the analytical solution given by Eq.~(\ref{secondordercorrelationfunction}) in Sec.~IV and numerical simulations based on master equation~({\ref{masterequation}}), respectively. The parameters chosen in Fig.~\ref{dip}(a) and (b) are the same as in Fig.~\ref{figure1}(a) and (b), respectively.}\label{dip}
 \end{figure}

\begin{figure}
  \centerline{
  \includegraphics[width=8.5cm, height=5.0cm, clip]{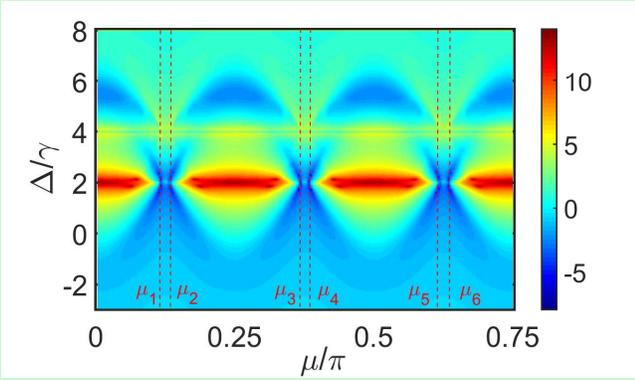}}
   \caption{The second-order correlation function $log_{10}[g^{(2)}(0)]$ of the CW mode as a function of $\Delta$ and $\mu$ is plotted by solving Eq.~(\ref{masterequation}). The parameters chosen are the same as in Fig.~\ref{figure1}(a). In this case, $\mu_j \, (j=1,2,\cdots,6)$ corresponding to EPs is the same as Fig.~\ref{figure31}. }\label{figure6}
 \end{figure}

To further clarify the equivalence between the effective Hamiltonian~(\ref{hamiltonian5}) and total Hamiltonian~(\ref{hamiltonian3}), we compare $g^{(2)}(0)$ with steady states of two cases for the fixed detuning $\Delta/\gamma= 2 $. The steady-state density operators are obtained from the numerical solutions of $ {d{\rho} }/{dt}=0$ in Eq.~(\ref{masterequation}) and $ {d{\rho}_{\rm{opt}} }/{dt}=0$ in Eq.~(\ref{master_equation_original}), respectively. Fig.~\ref{figure10}(a) displays $g^{(2)}(0)$ as a function of $\mu$ in the case of weak optomechanical coupling, where the numerical result corresponding to effective master equation ({\ref{masterequation}}) (blue-solid line) agrees well with that based on the total master equation ({\ref{master_equation_original}}) (red circles).
In addition, we also plot $g^{(2)}(0)$ as a function of $\Delta$ with various angles $\mu$, as shown in Fig.~\ref{figure10}(b), which suggests that the results of the effective master equation ({\ref{masterequation}}) and total master equation ({\ref{master_equation_original}}) are consistent for different $\mu$.

All of these calculations clearly show the equivalence of the optomechanical system given by Eq.~(\ref{master_equation_original}) and Kerr-type nonlinearity of Eq.~(\ref{masterequation}) in the appropriate parameter regime. It is valid that we use the effective master equation ({\ref{masterequation}}) instead of the total master equation ({\ref{master_equation_original}}) to solve this problem. In this case, we can reduce the dimension of the Hilbert space, which makes solving the problem much more accessible.

\section{\label{SupplementaryFigure} Supplementary discussion of Figure \ref{figure1}}

We note that there is a dip in $g^{(2)}(0) > 1$ for $\Delta/\gamma=4$ given by Fig.~\ref{figure1}(a) corresponding to the photon bunching, which is revealed from Fig.~\ref{dip}(a), where the dip appears in $g^{(2)}(0)$  for $\Delta/\gamma=4$ due to the fact that the cusp occurs in $P_{20}$.
Fig.~{\ref{dip}}(b) has the similar observation at $\Delta=6 \gamma$.

\begin{figure}[t]
\centerline{
\includegraphics[width=8.6cm, height=4.0cm, clip]{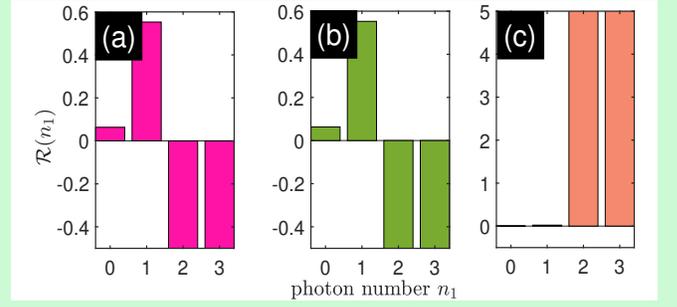}}
\caption{The figure shows the relative photon distributions $\mathcal{R}(n_1)=[P_{n_1}-\mathcal{P}_{n_1}]/\mathcal{P}_{n_1}$, i.e., the deviation of the photon distribution $P_{n_1}=\sum\nolimits_{n_2} {\langle {{n_1},{n_2}| \rho  |{n_1},{n_2}} \rangle }$ given by Eq.~(\ref{masterequation}) from the standard Poisson distribution $\mathcal{P}_{n_1} = {{\langle {\hat a_1^\dag {\hat a_1}} \rangle }^{n_1}}{e^{ - \langle {\hat a_1^\dag {\hat a_1}} \rangle}}/{{{n_1}!}}$ with the same photon number $n_1$ in the CW mode. The parameters chosen are (a) $\mu=\mu_1=0.1171\pi$ (position of the EP in Fig.~\ref{figure31}); (b) $\mu=\mu_2=0.1329\pi$ (position of the EP in Fig.~\ref{figure31}); (c) $\mu=0.125\pi$ (position of the non-EP). The other parameters are the same as Fig.~\ref{figure5}(a).}\label{figure7}
\end{figure}
\section{\label{SupplementaryCPB} Supplementary discussion of CPB}
The Appendix gives more discussions on the photon statistical properties of CPB at EPs. We first plot $g^{(2)}(0)$ of the CW mode in a logarithmic scale as a function of ${\Delta}$ and $\mu$, as shown in Fig.~\ref{figure6}. $g^{(2)}(0)$ changes periodically with the increase of $\mu$, which originates from the period of coupling coefficients $\mathcal {E}_{1}$ and $\mathcal {E}_{2}$ given by Eq.~(\ref{nonreciprocal}). It is worth noting that there are local minimum values of $g^{(2)}(0)$ at the EP $\mu_j$ when $\Delta$ is equal to $U$, i.e., ${\Delta/\gamma}=2$, as shown the intersections of horizontal (${\Delta}=2 \gamma$) and vertical ($\mu=\mu_j$) coordinates in Fig.~\ref{figure6}, which agrees well with the results in Fig.~\ref{figure1}(a). However, if we adjust the relative angular position $\mu$ away from EPs (e.g., $\mu=0.25\pi$), $g^{(2)}(0)$ arrives at the maximum for ${\Delta/\gamma}=2$.

Moreover, CPB at EPs can also be confirmed by comparing the photon-number distribution $P_{n_1}$ with the standard Poisson distribution $\mathcal{P}_{n_1}$. Therefore, we investigate the relative photon distributions of CW mode $\mathcal{R}({n_1})=[P_{n_1}-\mathcal{P}_{n_1}]/\mathcal{P}_{n_1}$, as shown in Fig.~\ref{figure7}. When we adjust $\mu$ to EPs ($\mu_1=0.1171\pi$ in Fig.~\ref{figure7}(a) and $\mu_2=0.1329\pi$ in Fig.~\ref{figure7}(b)), the photon-number distribution $P_1$ is greater than the standard Poisson distribution $\mathcal{P}_1$ for the single-excitation subspace (${n_1}=1$), i.e., $\mathcal{R}(1)=[P_1-\mathcal{P}_1]/\mathcal{P}_1>0$.
However, if the photon number ${n_1}$ is greater than or equal to $2$, the photon-number distribution $P_{n_1}$ $({n_1} \ge 2)$ is less than the standard Poisson distribution $\mathcal{P}_{n_1}$ $({n_1}\ge 2)$, i.e., $\mathcal{R}({n_1})<0$ $({n_1}\ge 2)$. It suggests that the photon is more inclined to exist singly, namely the sub-Poisson distribution, which is the antibunching effect. This confirms CPB at EPs from another aspect. If we tune $\mu$ away from EPs, e.g., $\mu=0.125\pi$ in Fig.~\ref{figure7}(c), the phenomenon is strikingly different, which indicates the bunching effect. In summary, Fig.~\ref{figure7} shows that $P_1$ is enhanced while $P_{n_1}$ $({n_1} > 1)$ is suppressed at EPs, which is in sharp contrast to the case where we change the relative angular position $\mu$ away from EPs.

\end{document}